\documentclass[a4paper,12pt]{article}
\pdfoutput=1
\usepackage{feynmp-auto,expdlist}
\usepackage{amsmath, amsfonts, amssymb}
\usepackage{graphicx}
\usepackage{enumerate}
\usepackage{hyperref}
\usepackage{latexsym}
\usepackage{dsfont}
\usepackage{hepnicenames}
\usepackage{enumerate}
\usepackage{soul}
\usepackage[normalem]{ulem}
\usepackage{bbold}

\usepackage{units}

\newcommand{\B}{{\cal B}}

\usepackage{mathrsfs,graphicx,rotating,amsmath,amsfonts,mathtools,booktabs,amssymb,wasysym}
\usepackage{hyperref}\usepackage{slashed}
\usepackage[nosort]{cite}
\usepackage[table,xcdraw,dvipsnames]{xcolor}
\usepackage{bm}
\usepackage{graphicx}
\usepackage{multirow,multicol}
\hypersetup{colorlinks,bookmarksopen,bookmarksnumbered,
	linkcolor=blus,pdfstartview=FitH,urlcolor=rossos,citecolor=verde}
\allowdisplaybreaks

\newcommand{\tauDG}{\tau_{\scriptscriptstyle{\rm DG}}}

\newcommand{\GammaDG}{\Gamma_{\scriptscriptstyle{\rm DG}}}
\renewcommand{\[}{\left[}
\renewcommand{\]}{\right]}
\renewcommand{\(}{\left(}
\renewcommand{\)}{\right)}
\renewcommand{\S}{{\cal S}}

\newcommand{\sigmav}{\langle \sigma v \rangle}

\newcommand{\mio}[1]{}

\def\bpm{\begin{pmatrix}}
	\def\epm{\end{pmatrix}}

\usepackage{mathrsfs}

\newcommand{\fig}[1]{~\ref{fig:#1}}
\newcommand{\sfrac}[2]{#1/#2}
\newcommand{\One}{1\!\!\hbox{I}}

\allowdisplaybreaks
\usepackage{multicol}
\usepackage{color}
\definecolor{rosso}{cmyk}{0,1,1,0.4}
\definecolor{rossos}{cmyk}{0,1,1,0.55}
\definecolor{rossoc}{cmyk}{0,1,1,0.2}
\definecolor{blu}{cmyk}{1,1,0,0.3}
\definecolor{blus}{cmyk}{1,1,0,0.6}
\definecolor{bluc}{cmyk}{1,1,0,0.1}
\definecolor{verde}{cmyk}{0.92,0,0.59,0.25}
\definecolor{verdec}{cmyk}{0.92,0,0.59,0.15}
\definecolor{verdes}{cmyk}{0.92,0,0.59,0.4}

\newcommand{\Tr}{\,{\rm Tr}}
\newcommand{\diag}{\,{\rm diag}}

\def\circa#1{\,\raise.3ex\hbox{$#1$\kern-.75em\lower1ex\hbox{$\sim$}}\,}

\newcommand{\beq}{\begin{equation}}
\newcommand{\eeq}{\end{equation}}

\newcommand{\bea}{\begin{eqnarray}}
\newcommand{\eea}{\end{eqnarray}}
\newcommand{\be}{\begin{equation}}
\newcommand{\ee}{\end{equation}}
\font\tenrsfs=rsfs10 at 12pt
\font\sevenrsfs=rsfs7
\font\fiversfs=rsfs5
\newfam\rsfsfam
\textfont\rsfsfam=\tenrsfs
\scriptfont\rsfsfam=\sevenrsfs
\scriptscriptfont\rsfsfam=\fiversfs

\newcommand{\D}{{\cal D}}

\newcommand{\La}{\mathscr{L}}

\newcommand{\I}{\mathbb{1}}

\newsavebox\MBox

\newcommand{\Sp}{\,{\rm Sp}}

\newcommand{\SU}{\,{\rm SU}}
\newcommand{\SO}{\,{\rm SO}}
\newcommand{\U}{\,{\rm U}}
\renewcommand{\L}{\mathscr{L}}

\def\circa#1{\,\raise.3ex\hbox{$#1$\kern-.75em\lower1ex\hbox{$\sim$}}\,}
\makeatletter

\font\ital=cmu10

\def\hhref#1{\href{http://arxiv.org/abs/#1}{arXiv:#1}}
\usepackage{xstring}
\newcommand{\hhrefq}[1]{\IfSubStr{#1}{:}{\href{http://inspirehep.net/search?ln=en&ln=en&p=#1&of=hb&action_search=Search&sf=&so=d&rm=&rg=25&sc=0}{InSpire:#1}}{\hhref{#1}}}

\def\art{\@ifnextchar[{\eart}{\oart}}
\def\eart[#1]#2#3#4#5#6{{\rm #2}, {\em #3 \bf #4} {\rm (#6) #5} ({\em #1})}
\def\article{\@ifnextchar[{\earticle}{\oarticle}}
\def\oarticle#1#2#3#4#5#6{{\rm #1}, {\ital ``#6''}, {\rm #2 #3 (#5) #4}}
\def\earticle[#1]#2#3#4#5#6#7{{\rm #2}, {\ital ``#7''}, {\rm #3 #4 (#6) #5}  [\hhrefq{#1}]}
\def\hepart[#1]#2{{\rm #2, \sl#1}}
\def\heparticle[#1]#2#3{#2, {\ital ``#3''} [\hhrefq{#1}]}
\newcommand{\doi}[1]{\href{http://dx.doi.org/#1}{[link]}}

\newcommand{\hhrefqq}[1]{\IfBeginWith{#1}{10.}{\href{https://doi.org/#1}{doi:#1}}{\hhrefq{#1}}}
\def\earticle[#1]#2#3#4#5#6#7{{\rm #2}, {\ital ``#7''}, {\rm #3 #4 (#6) #5}  [\hhrefqq{#1}]}

\newcounter{alphaequation}[equation]
\def\thealphaequation{\theequation\hbox to
	0.6em{\hfil\alph{alphaequation}\hfil}}
\def\eqnsystem#1{
	\def\@eqnnum{{\rm (\thealphaequation)}}
	\def\@@eqncr{\let\@tempa\relax \ifcase\@eqcnt \def\@tempa{& & &} \or
		\def\@tempa{& &}\or \def\@tempa{&}\fi\@tempa
		\if@eqnsw\@eqnnum\refstepcounter{alphaequation}\fi
		\global\@eqnswtrue\global\@eqcnt=0\cr}
	\refstepcounter{equation} \let\@currentlabel\theequation \def\@tempb{#1}
	\ifx\@tempb\empty\else\label{#1}\fi
	\refstepcounter{alphaequation}
	\let\@currentlabel\thealphaequation
	\global\@eqnswtrue\global\@eqcnt=0 \tabskip\@centering\let\\=\@eqncr
	$$\halign to \displaywidth\bgroup \@eqnsel\hskip\@centering
	$\displaystyle\tabskip\z@{##}$&\global\@eqcnt\@ne
	\hskip2\arraycolsep\hfil${##}$\hfil& \global\@eqcnt\tw@\hskip2\arraycolsep
	$\displaystyle\tabskip\z@{##}$\hfil
	\tabskip\@centering&\llap{##}\tabskip\z@\cr}
\def\endeqnsystem{\@@eqncr\egroup$$\global\@ignoretrue} \makeatother

\definecolor{Gray}{gray}{0.95}

\def\bal#1\eal{\begin{align}#1\end{align}}

\newcommand{\W}{\mathcal{W}}
\newcommand{\Z}{\mathcal{Z}}
\newcommand{\X}{\mathcal{X}}
\newcommand{\A}{\mathcal{A}}

\newcommand{\N}{\mathcal{N}}
\newcommand{\s}{\mathfrak{s}}
\newcommand{\g}{\mathfrak{g}}
\newcommand{\M}{\mathcal{M}}
\newcommand{\G}{\mathcal{G}}
\renewcommand{\H}{\mathcal{H}}

\begin{document}
\vspace{1.5cm}

\begin{center}
	{\Large\LARGE\Huge \bf \color{rossos} Dark Matter in scalar $\Sp(\N)$ gauge dynamics}\\[1cm]
	{\bf Giacomo Landini$^{a,b}$,
	 Jin-Wei Wang$^{b,c,d}$}\\[7mm]
	
	{\it $^a$ INFN, Sezione di Pisa, Italy}\\[1mm]
	{\it $^b$ Dipartimento di Fisica dell'Universit{\`a} di Pisa}\\[1mm]
	{\it $^c$ Key Laboratory of Particle Astrophysics, Institute of High Energy Physics,
		Chinese Academy of Sciences, Beijing, China}\\[1mm]
	{\it $^d$ School of Physical Sciences, University of Chinese Academy of Sciences, Beijing, China}\\[1mm]
	\vspace{0.5cm}
	{\large\bf\color{blus}}
	
	\begin{quote}\large
	We consider a model  with Sp dark gauge group and a scalar field in the fundamental representation, which leads to two co-stable DM candidates at the perturbative level thanks to a global $\U(1)$ accidental symmetry. After gauge confinement at low energy scale, only one of the two candidates is still stable.
	We compute the DM relic abundance by 
	solving the Boltzmann equations numerically. 
	The presence of light dark glueballs gives extra cosmological effects and can affect Higgs physics. We study the DM phenomenology, providing the predictions for direct and indirect detection (including the Sommerfeld enhancement). We show that the model predicts a slightly suppressed indirect detection cross section in comparison to the usual WIMPs paradigm.
	\end{quote}
	
	\thispagestyle{empty}
	\bigskip
	
\end{center}

\newpage

\setcounter{footnote}{0}

\tableofcontents

\section{Introduction}\label{intro}
The possibility that Dark Matter (DM) originates from elementary scalar/gauge dynamics has been explored in \cite{Buttazzo:2019iwr} and \cite{Buttazzo:2019mvl}.
In particular, \cite{Buttazzo:2019iwr} considered one new elementary scalar field $\S$ that fills the fundamental representation of a dark gauge group $\G = \{\SU(\N),\SO(\N),\Sp(\N),G_2\}$. Every choice of $\G$ is characterized by an accidental symmetry, which leads to stable DM candidates with non-trivial features that are characteristics of each group. DM is always accompanied by light dark glueballs.
	
The dark group $\G$ can become strongly interacting ('confined phase') and/or get spontaneously broken by vacuum expectation values of $\S$ (‘Higgsed phase’ breaking $\G$ to a subgroup $\H$ that will also confine at low energy scale). A surprising equivalence between the confined
and Higgsed phases for scalars $\S$ in the fundamental representation of the gauge groups $\G$ has been studied in~\cite{Buttazzo:2019iwr,Hambye:2009fg}. In these models $\S$ has an unique self-quartic, that leads to an
unique symmetry breaking pattern where the only surviving scalar is a Higgs-like singlet, named scalon.
	
It is interesting to study the DM phenomenology of these models. Since the Higgs and confined phases, thanks to the equivalence between them, contain the same asymptotic particles, one can perform all computations in the Higgs phase for weak couplings. The phenomenology can be very different for each gauge group. In the $\SU$ case there turns out to be two DM candidates but one of them gives a negligible contribution to the DM abundance. The related phenomenology has been studied in detail in \cite{Buttazzo:2019iwr}. The situation for the $\Sp$ dynamics is more involved: at the perturbative level there are two co-stable DM candidates arising from the spontaneous symmetry breaking (SSB)  $\Sp(\N+2)\to\Sp(\N)$. After the confinement of the subgroup $\Sp(\N)$, one of two candidates forms bound states (mesons) that can decay, leaving only one DM candidate. 
Therefore, we expect the DM phenomenology for $\Sp$ case to be more intriguing and worthy of more detailed calculations.
	
In this paper we compute the DM relic abundance of the Sp DM model: we consider a $\Sp$ dark gauge group and a scalar field that fills the fundamental representation and we solve the related Boltzmann equations numerically. We make some simplified assumptions to take into account gauge confinement, and we also consider the presence of light glueballs, which can dilute the DM relic abundance.

It is well known that light mediators can affect DM annihilations cross section through an enhancement factor (Sommerfeld enhancement) and bound state formation~\cite{Mitridate:2017izz,Oncala:2018bvl}. The $\Sp$ DM model includes two kinds of light mediators: a scalar particle (the scalon) and massless gauge bosons, which arise from the spontaneous symmetry breaking of the gauge group. 
We also introduce how to compute Sommerfeld and bound state corrections and how to take them into account in the predictions for both the relic abundance and indirect detection signals.
	
We calculate the relevant cross sections for the direct and indirect detections and compare them with the current experimental limits as well as the future prospects, and we find that the Sp DM model is testable on the future indirect detection experiment like CTA~\cite{Morselli:2017ree}. We also investigate the phenomenology of the scalon and of light glueballs and their connection with Higgs physics.
	
This paper is outlined as follows: in section \ref{modelR} we introduce the main features of our model, including the Lagrangian, the breaking pattern, the perturbative mass spectrum, the accidental symmetries and the formation of bound states after confinement.
In section \ref{Sommerfeld} we discuss the Sommerfeld enhancement factors for annihilation cross section and  the formation of perturbative bound states.
In section \ref{DMrelic} we compute the DM relic abundance by solving the Boltzmann equations. In section \ref{pheno1} we discuss the predictions and the experimental limitations of this model, including scalon production, direct and indirect detections. In section \ref{sec:scale_inv} we briefly analyse the effects of the quadratic term in the scalar potential on the spectrum and the phenomenological results by abandoning the scale invariant hypothesis.  Conlusions and the summary of results are given in section \ref{conclusions}. Appendix~\ref{app:SpNFeyn} gives the explicit expressions of the Feynman rules of the Sp DM model in the Higgs phase.

\section{Model detail}\label{modelR}
\subsection{Lagrangian and mass spectrum}\label{model}
The group $\Sp(\N)$ is defined for even $\N$  
as the transformations that leave
invariant the tensor $\gamma_\N\equiv \One_{\N/2}\otimes \epsilon$, where
$\epsilon_{ij}$ is the 2-dimensional anti-symmetric tensor. We consider a new $\Sp(\N+2)$ gauge sector, with vectors ${\G}^a_{\mu\nu}$ in the adjoint and one scalar field $\S$ in the fundamental representation,
which is pseudo-real\footnote{We remind the reader that the $\Sp(\N)$ generators can be written as
\begin{equation*}
\frac{T_{\rm asym} \otimes \I_2}{\sqrt{2}}, \qquad \frac{T_{\rm sym} \otimes \sigma_k}{\sqrt{2}},
\end{equation*}
where $\sigma_k$ ($k=1,2,3$) are the Pauli matrices, $T_{\rm sym} = \{T_{\rm real}, \I_{\N/2}/\sqrt{\N}\}$ are the symmetric generators of $\U(\N/2)$, and $T_{\rm asym} = T_{\rm imag}$ are the anti-symmetric ones\cite{Buttazzo:2019mvl}.}.
We thereby introduce a complex $(\N+2)$-dimensional scalar $\S$.
The Lagrangian is given by
\beq
\La=\La_{\rm SM}-\frac{1}{4} {\G}^a_{\mu\nu}{\G}^{a\,\mu\nu} -V_\S+	|\D_\mu\S|^2,
\label{eq:la1}
\eeq
with scalar potential
\begin{equation} \label{pot}
  V_\S =
-M_\S^2 |\S|^2 +  \lambda_\S |\S|^4 -  \lambda_{H\S} |H|^2 |\S|^2,
\end{equation}
where $H$ is the SM Higgs field.
The Lagrangian conserves an accidental global U(1) symmetry (under which $\S$ has charge 1),
by virtue of $\S^T \gamma \S=0$. The RGEs at one-loop level are
\begin{eqnsystem}{sys:RGESOn}
	(4\pi)^2 \frac{d\g}{d\ln\mu} & = & -\frac{11(\N+2)+21}{6} \g^3\,,\\
	(4\pi)^2 \frac{d\lambda_\S}{d\ln\mu} &=&\frac{3}{16} (\N+6) \g^4 -
	3(\N+3) \g^2 \lambda _\S +4(6+\N)  \lambda^{ 2} _\S,
\end{eqnsystem}
where $\g$ is the gauge coupling of the new sector. We normalize the generators in the fundamental representation as $\Tr[T^a,T^b]=\frac{1}{2}\delta^{ab}$. It is interesting to consider dynamical symmetry breaking through the Coleman-Weinberg mechanism, obtained by setting $M_\S=0$ (scale invariant hypothesis). In section \ref{sec:scale_inv} we briefly discuss the effects of the quadratic term $M_\S^2$ by abandoning the scale invariant hypothesis. Assuming that $\lambda_{H\S}$ is negligibly small, the RGEs give that $\S$ can radiatively acquire a vacuum expectation value
$\S(x)=(0,...,0,(w+\s(x))/\sqrt{2})^T$, breaking the gauge group $\Sp(\N+2)$ to $\Sp(\N)$. 
At the same time, the accidental U(1) symmetry gets rotated
to an unbroken global U(1) with generator 
\beq \label{eq:TdiagSp}
\diag(1,\ldots,1,1,1) + \diag(0,\ldots,0,1,-1) =  \diag (1,\ldots,1 ,2,0) ,\eeq
obtained by combining the original U(1) with the diagonal generator of the broken $\Sp(2)$. We call this unbroken global $\U(1)$ dark-baryon number. If the Higgs mass term is absten too, this model can also generate the weak scale $v\simeq246$ GeV as~\cite{Hambye:2013sna}
\begin{equation}\label{vev}
v\simeq w\sqrt{\frac{\lambda_{H\S}}{2\lambda_H}},
\end{equation}
assuming a small positive $\lambda_{H\S}$ (more complicated expressions hold if $\lambda_{H\S}$ is not negligibly small).

Writing the gauge bosons as 
\beq 
T^a \G^{a}_\mu = 
\left(
\begin{array}{c|cc}
	\A_\mu & \X^{\dagger}_\mu/2  & \gamma_{\N}\X_\mu/2 \\ \hline
	\X_\mu/2 & \Z_\mu/2 &  \W_\mu/\sqrt{2} \\
	\gamma_{\N}\X^{\dagger}_\mu/2 & - \W_\mu^{\dagger}/\sqrt{2} & -\Z_\mu/2
\end{array}
\right),
\eeq
the perturbative mass spectrum in the broken phase is:
\begin{itemize}
	\item the singlet scalon $\s$ with one-loop supressed mass $M_\s^2=\beta_{\lambda_\S}w^2$;
	\item one real $\Z$ with mass $M_\Z = \g w/2$ and zero dark-baryon number;
	\item one complex $\W$, with mass $M_\W=M_\Z$ and dark-baryon number 2.
	\listpart{For $\N=0$ this is the $\Sp(2)=\SU(2)$ model of~\cite{Hambye:2008bq,Hambye:2009fg,Hambye:2013sna}
		where $\W$ and
		$\Z$ are co-stable DM candidates.  For $\N\ge 2$ the spectrum contains extra particles:}
	\item  $\N$ complex massive vectors $\X$ in the fundamental representation of $\Sp(\N)$
	with mass $M_{\X}=M_{\W}/\sqrt{2}$ and dark-baryon number 1;
	\item $\N(\N+1)/2$ massless vectors $\A$ in the adjoint representation of $\Sp(\N)$.
\end{itemize}
The $\Z$ boson decays into $\A$'s through loops involving the $\Z \W \W^{\dagger}$ and $\Z \X \X^{\dagger}$  gauge couplings. At perturbative level the $\W$ and $\X$ are DM candidates, co-stable  thanks to accidental baryon  number conservation.

\subsubsection*{Condensation of $\Sp(\N)$}
The theory becomes strongly coupled and the unbroken $\Sp(\N)$ confines at the energy scale
\begin{equation}\label{lambdaDC}
\Lambda_\text{DC}=M_\W \text{exp}\left[-\frac{12\pi}{11(\N+2)\alpha(M_\W)}\right].
\end{equation}
with $\alpha(M_\W)=\g^2(M_\W)/4\pi$, where $\g^2(M_\W)$ is the value of the dark gauge coupling at the scale $M_\W$.
After confinement we get the following spectrum of 
asymptotic states:
\begin{itemize}
	\item The scalon $s$, the $\Z$ and $\W$ bosons, and dark glue-balls $\A\A$.
	\item Two kinds of dark mesons: the unstable $\X^\dagger\X$ and $\X^\dagger \D_{\mu} \X$, which have the same quantum numbers as $s$ and $\Z$, and $\M_\mu=\X^T \gamma_{\N} \D_\mu\X$,
	with dark-baryon number 2 as $\W$. Only one linear combination of $\M$ and $\W$ appears among the stable asymptotic states, while the other corresponds to a resonance. 
	A similar situation holds for $s$ and $\X^\dag\X$, and for $\Z$ and $\X^\dag\D\X$.
	
	\item Dark baryons $\B$
	(defined as states formed with one $\epsilon_{i_1\cdots i_{\N-2}}$ tensor)
	are not stable because the $\epsilon$ tensor can be decomposed as
	$\epsilon_{i_1\cdots i_{\N-2}} = \gamma_{i_1 i_2}\cdots \gamma_{i_{\N-3}i_{\N-2}} + \hbox{permutations}$~\cite{Witten:1983tx}. 
	This means that $\B$ splits into $\N/2-1$ mesons $\M$.
\end{itemize}
Both the $\W$ and the mesons $\M$  carry charge 2 under conserved U(1) baryon number.

\subsection{Lagrangian in Higgs phase: $\Sp(\N+2)$$\rightarrow$$\Sp(\N)$}
After symmetry breaking the Lagrangian in eq.\eqref{eq:la1} can be further expanded as:
\begin{equation}\label{Lag}
-\frac{\mathcal{G}^a_{\mu\nu}\mathcal{G}^{a\mu\nu}}{4}+|D_\mu\mathcal{S}|^2= \L_{\text{kin}}+\L_{{\W\Z}}+\L_{{\W\X}}+\L_{{\X\Z}}+\L_\text{scalar}+\L_{\text{other}}.
\end{equation}
The kinetic terms are
\begin{equation}
\L_{\text{kin}}=-\frac{1}{4}\mathcal{A}_{\mu\nu}^a\mathcal{A}^{a\mu\nu}-\frac{1}{2}\mathcal{W}_{\mu\nu}^{\dagger}\W^{\mu\nu}-\frac{1}{4}\Z_{\mu\nu}\Z^{\mu\nu}-\frac{1}{2}\mathcal{X}_{\mu\nu}^{\dagger}\X^{\mu\nu}.
\label{eq:f1}
\end{equation}
The interaction terms between $\W\Z$ and $\W\X$ are
\begin{align}
\L_{{\W\Z}}=&+ i\g  \((\partial_{[\mu} \W_{\nu]}^\dag)\W^\nu \Z^\mu - \W_\nu^\dag (\partial^{[\mu}\W^{\nu]})\Z_\mu - \W_{[\mu}^\dag \W_{\nu]} \partial^\mu \Z^\nu)\)\notag\\
&-{\g^2} \( \W^\dag_\mu \W^\mu \Z_\nu \Z^\nu - \W^\dag_\mu \W_\nu \Z^\mu \Z^\nu \),
\end{align}
\begin{align}
\L_{{\W\X}}=&-\frac{\g^2}{2} \(\W^\dag_\mu\W^\mu\X^\dag_\nu\X^\nu+\W^\dag_\mu\W_\nu\X^{\dag\nu}\X^\mu-2\W^\dag_\mu\W_\nu\X^{\dag\mu}\X^\nu\)\notag\\
&- \frac{i}{\sqrt{2}}\g  \(\frac{1}{2}\gamma_{\N}^{ij}(\X^i_{[\nu} \X^j_{\mu]})\partial^\mu \W^{\dagger\nu}+\gamma_{\N}^{ij}\W^\dagger_{[\nu}\X^i_{\mu]}\partial^\mu\X^{j\nu}\).
\end{align}
The interaction terms between $\X\Z$ are
\begin{align}
\L_{{\X\Z}}=&\frac{i}{2}\g  \((\partial_{[\mu} \X_{\nu]}^\dag)\X^\nu \Z^\mu - \X_\nu^\dag (\partial^{[\mu}\X^{\nu]})\Z_\mu - \X_{[\mu}^\dag \X_{\nu]} \partial^\mu \Z^\nu)\)\notag\\
&-\frac{\g^2}{4} \(\X^\dag_\mu\X^\mu\Z^\nu\Z_\nu-\X^\dag_\mu\X_\nu\Z^{\mu}\Z^\nu\).
\end{align}
The kinetic and interaction terms for the scalar are 
\begin{align}
\L_\text{scalar}=&\frac{1}{2} \partial_\mu \s \partial^\mu \s + M^2_\W (1+ \frac{\s}{w})^2 \W^\dag_\mu \W^\mu
+ \frac{1}{2} M^2_\Z (1+ \frac{\s}{w})^2 \Z_\mu \Z^\mu \notag\\
&+M^2_\X (1+ \frac{\s}{w})^2 \X^\dag_\mu\X^\mu.
\end{align}
The other interaction terms between $\W$, $\Z$, $\X$ and $\A$ are
\begin{align}
\L_{\text{other}}=&-i\g \(\X^\dag_\mu T^a_{\N}\X_\nu\)\A^{a\mu\nu}\notag\\
&-\frac{\g^2}{2} \(-\(\Z_\mu\A_\nu^a+\Z_\nu\A_\mu^a\)\(\X^{\dagger\mu}T^a_{\N}\X^\nu\)+2\Z_\mu\A^{\mu a}\(\X^{\dagger}_{\nu}T^a_{\N}\X^\nu\)\)\notag\\
&-\frac{3\g^2}{8\sqrt{2}}\gamma_{\N}^{ij}\X^i_{[\mu}\X^j_{\nu]}\(\W^{\dagger\nu}\Z^\mu-\W^{\dagger\mu}Z^\nu\)\notag\\
&+\frac{\g^2}{\sqrt{2}}\gamma_{\N}^{il}T^a_{lk}\left(\X_\mu^i\X^{\mu k}\W^\dag_\nu\A^{a\nu}-\frac{1}{2}\X_{\{\mu}^i\X_{\nu\}}^k\W^\dag_\mu\A^{a\nu}\right)+\cdots,
\label{eq:f2}
\end{align}
where $\cdots$ denotes $\W\W^{\dagger}\W\W^{\dagger}$ and $\X\X^{\dagger}\X\X^{\dagger}$  interactions and
\begin{equation*}
\W_{\mu\nu} = \partial_\mu \W_\nu - \partial_\nu \W_\mu,\qquad
\Z_{\mu\nu} = \partial_\mu \Z_\nu - \partial_\nu \Z_\mu,\qquad
\X_{\mu\nu} = \D_\mu \X_\nu - \D_\nu \X_\mu,
\end{equation*}
\begin{equation*}
\A^a_{\mu\nu} = \partial_\mu \A^a_\nu - \partial_\nu \A^a_\mu + \g f^{abc}_{\N} \A^b_\mu \A^c_\nu,\qquad \D_\mu=\partial_\mu-i\g T^a_{\N}\A^a_\mu;
\end{equation*}
$f^{abc}_{\N}$ are the structure constants and $T^a_{\N}$ are the generators of the $\Sp(\N)$ group in the fundamental representation.
Schematically, the interaction vertices can be summarized as:
	\begin{itemize}
		\item 3-gauge vertices: $\A\A\A$, $\A\X\X^\dagger$, $\X\X\W^\dagger$, $\Z\W\W^\dagger$, $\Z\X\X^\dagger$;
	
		\item 4-gauge vertices: $\A\A\A\A$, $\A^2\X\X^\dagger$, $\A\Z\X\X^\dagger$, $\A\X^2\W^\dagger$, $\Z\X^2\W^\dagger$, $\Z^2\W\W^\dagger$, $\Z^2\X\X^\dagger$, \\ $\X\X^\dagger\W\W^\dagger$, $(\W\W^\dagger)^2$, $(\X\X^\dagger)^2$;
		
		\item Gauge-scalar vertices: $\s\Z\Z$, $\s\s\Z\Z$, $\s\W\W^\dagger$, $\s\s\W\W^\dagger$,$\s\X\X^\dagger$, $\s\s\X\X^\dagger$.
	\end{itemize}
By using the eq.\eqref{eq:f1}$\sim$eq.\eqref{eq:f2} we can get the desired  Feynman rules for the interaction vertices, which are collected in Appendix \ref{app:SpNFeyn}.

\section{Sommerfeld enhancement and bound states}\label{Sommerfeld}
The DM relic abundance as well as indirect detection signals are determined by the non-relativistic annihilation (and co-annihilation) cross sections among DM particles and with every particle that can interact with them. By using the Feynman rules listed in Appendix \ref{app:SpNFeyn}, we compute explicitly these cross sections in section \ref{crosssections}. Besides, the DM relic abundance and the indirect detection predictions are calculated in section \ref{relic} and section \ref{indirectsection}, respectively.

The annihilation cross sections get enhanced if the particles in the intial states can exchange a light mediator, a contribution known in the literature as Sommerfeld enhancement~\cite{andp.19314030302,Sakharov:1991pia,Hisano:2006nn,Cirelli:2007xd,Belotsky:2005dk}. This effect can be relevant for the computation of DM relic abundance as well as (and especially) for indirect detection. Generally speaking, it can be taken into consideration by multiplying the cross sections by an enhancement factor $S$.

In this section we summarize how to compute the enhancement factor for all the interesting cases that occur in our model. All the interaction processes that we are interested in can be divided into two categories: 
\begin{itemize}
	\item $\X\X^{\dagger}$ and $\X\X$ annihilations: they can exchange a massless dark gluon $\A$ generating a Coulomb potential. If the state $\X\X^{\dagger}$($\X\X$) is in the representation $J$ of $\Sp(\N)$, the cross section gets enhanced by~\cite{Mitridate:2017izz}:
	\begin{equation}\label{SA}
	S_{\A}=\frac{2\pi \alpha_{\text{eff}}/v_\text{rel}}{1-e^{-2\pi\alpha_{\text{eff}}/v_\text{rel}}}
	\end{equation}
	with $\alpha_{\text{eff}}=\lambda_J\alpha$, where $\alpha=\g^2/4\pi$ and $\lambda_J$ is a group theory factor which is computed, as explained in~\cite{Mitridate:2017izz}, as $\lambda_J=C_R-\sfrac{C_J}{2}$, where $C_J$ and $C_R$ are the Casimir for the $J$ and fundamental reresentation, respectively. $v_\text{rel}$ is relative velocity among two particles $\X$.

	
	\item all other processes: the two particles in the intial state are \{$\W\W^{\dagger}$, $\W\Z$, $\W\X^{\dagger}$, $\Z\Z$\}. For every case, they can exchange a scalon (with mass $M_s$) generating an attractive potential. In the non-relativistic limit, the potential can be expressed as:
	\begin{equation}\label{potential}
	V(r)=-\frac{\g^2}{16\pi}\frac{e^{-M_\s r}}{r},
	\end{equation}
	so the cross sections get enhanced by~\cite{Mitridate:2017izz}:
	\begin{equation}
	S_\s(M)=\frac{2\pi\alpha_\text{eff}}{v_\text{rel}}\frac{\sinh(\pi M v_\text{rel}/\kappa M_\s)}{\cosh(\pi M v_\text{rel}/\kappa M_\s)-\cosh(\pi M v_\text{rel}\sqrt{1-4\alpha_\text{eff}\kappa M_\s/M v_\text{rel}^2}/\kappa M_\s)},
	\label{eq:sommer1}
	\end{equation}
\end{itemize}
where $\kappa \approx 1.74$, $\alpha_\text{eff} = \g^2/16 \pi$, $v_\text{rel}$ is relative velocity among two particles of mass $M$. Under the thermal equilibrium assumption, $v_\text{rel}$ can be estimated as $v_\text{rel}=\sqrt{2}v=\sqrt{\sfrac{16T}{\pi M}}$ (if the two particles have different masses (i.e. $M_1$ and $M_2$), one can use $v_\text{rel}=\sqrt{8T(M_1+M_2)/\pi M_1 M_2}$).
\newline

There is a second phenomenon, which is related to Sommerfeld enhancement, can affect both the relic abundance and the indirect detection cross section: two scattering particles can form bound states through the exchange of a light mediator~\cite{Mitridate:2017izz,Oncala:2018bvl,Wise:2014jva,vonHarling:2014kha,Petraki:2015hla,Ellis:2015vaa,Asadi:2016ybp,Liew:2016hqo,Cirelli:2016rnw}. In particular, the interesting process is the formation of bound states via emission of the corresponding mediator (if the mediator is too heavy one can consider processes with emission of lighter particles, such as light glueballs). 

The physics is very different for massive or massless mediators: if the mediator is massive, which is the case for the scalon exchange (with mass $M_\s$) among $\W$, $\X$, and $\Z$ that we mentioned above, bound states can exist only if they satisfy the condition $\alpha_{\text{eff}}M>\kappa M_\s$ (for the ground state; for excited states the condition is even stronger)~\cite{Mitridate:2017izz,Oncala:2018bvl}, where $\alpha_\text{eff} = \g^2/16 \pi$ and $M=\{M_\W,M_\X\}$ is the mass of the particles involved in the scattering. However, this condition is never satisfied in our model,  so there will be no bound states generated by potential (16).

If instead the mediator is massless, which is the case for dark gluons exchange between $\X\X^{\dagger}$($\X\X$), the corresponding potential is Coulombian, and therefore bound states are always allowed. In particular one gets the usual infinity of bound states with binding energies $E_n=\alpha_{\text{eff}}^2 M_\X/4n^2$, with $\alpha_{\text{eff}}$ as in eq.\eqref{SA}.


\section{DM relic abundance}\label{DMrelic}
In this section we compute the cosmological relic abundance of DM by solving the related Boltzmann equations. As first step we consider perturbative annihilations and semi-annihilations of co-stable $\W$ and $\X$, which occur at high temperatures. Then we take into account the non-perturbative dynamics effects, i.e. the confinement of the $\Sp(\N)$ gauge group and the corresponding hadronization that become important at lower temperatures, in section \ref{relicsection}.

We recall that for $\N=0$ our model coincides with the $\Sp(2)=\SU(2)$ model, which has already been studied in~\cite{Hambye:2008bq,Hambye:2009fg,Hambye:2013sna}. Therefore, we will focus on the original cases for $\N\geq2$.
\subsection{Boltzmann Equations}
\subsubsection*{Notations}
We use the following notations:
\begin{itemize}
	\item we define a dimensionless parameter $z=\sfrac{M_\X}{T}$, where $T$ is the temperature of the universe;
	\item$H=(\sfrac{T^2}{M_{\text{P}}})\sqrt{\sfrac{4\pi^3g_{*s}}{45}}$ is the Hubble constant ($M_{\text{P}}$ is the Planck mass);
	\item $s=\sfrac{2\pi^2 g_{*s}T^3}{45}$ is the entropy density;
	\item $g_{*s}$ is the number of relativistic d.o.f. If freeze-out of DM happens before EW phase transition,
	$g_{*s}=106.75+\N(\N+1)+1$ where the extra d.o.f. comes from the massless dark gluons $\A$ and the presumibly light scalon $\s$;
	\item $Y_i=Y_i(z)=\sfrac{n_i}{s}$ is the number density  of the $i$ species divided by the entropy density;
	\item $Y_{i,\text{eq}}=Y_{i,\text{eq}}(z)$ is the equilibrium distribution of the $i$ species. Before DM thermal freeze-out, all species are in thermal equilibrium.
	At freeze-out all species except  DM ($\W$ and $\X$) keep staying in thermal equilibrium. DM is non-relativistic at freeze-out (which occurs roughly at $z\sim25$). We remind that the equilibrium distribution for a non-relativistic species is given by $Y_{i,\text{eq}}\simeq0.145 g_i \left(M_i/T\right)^{\sfrac{3}{2}}e^{-M_i/T}/g_{*s}$ where $M_i$ is the mass of the particle and $g_i$ its number of d.o.f. ($g_\W=6$, $g_\X=6\N$, $g_\Z=3$);
	\item $\sigmav(i\to j)$ is the thermal-averaged cross section for the $i\to j$ process in the non-relativistic limit; assuming s-wave annihilations we can take $\sigmav\simeq\sigma_0$.
\end{itemize}

\subsubsection*{Numerical factors}
We use the following convention in computing the cross sections and deriving the Boltzmann equations:
\begin{itemize}
	\item a factor 1/2 for identical particles in the final states is included in the cross sections;
	\item a factor $\kappa=1/2$ for complex DM annihilations is included in the cross sections (more precisely $\kappa=1/2$ if the inital states are both complex; $\kappa=1$ if the intial states are both real or one is real and one is complex);
	\item we put a factor 1/2 for identical particles in the initial state in the equations;
	\item we put a factor $\pm\Delta_i$ if in the process the number of DM particles of kind "$i$" increases (decreases) of $\Delta_i$ units.
\end{itemize}

With the above notations and numerical factors, we can write the general form of the Boltzmann equations as
\begin{equation}
\left(\frac{Hz}{s}\right)\frac{dY_i}{dz}=\sum_{\text{processes}}\Delta_i S \sigmav(12\to34)\left[Y_1Y_2-\frac{Y_3Y_4}{Y_3^{\text{eq}}Y_4^{\text{eq}}}Y_1^{\text{eq}}Y_2^{\text{eq}}\right],
\label{eq:be}
\end{equation}
where $S=1/2$ if the inital particles are identical and $S=1$ for other cases.
\subsubsection*{List of (semi-)annihilation processes}
All the kinematics permitted annihilation and semi-annihilation processes that can change the DM number of 1 or 2 units are listed below (Note that some processes, e.g. $\W\to\X\X$, are forbidden by energy-momentum conservation):
\begin{itemize}
	\item $\W^\dagger\W\rightarrow \X^\dagger\X, \s\s, \Z \s$;
	\item $\X^\dagger\X\rightarrow \A\A, \s\s, \Z \s, \A \s, \Z\A$;
	\item $\W\Z\rightarrow \X\X$;
	\item $\X^\dagger\W\rightarrow \A\X, \s\X$;
	\item $\X\X\rightarrow \A\W, \s\W$;
	\item $\Z\Z\rightarrow \X^\dagger\X$.
\end{itemize}
\subsubsection*{Equations}
By using the eq.\eqref{eq:be} and all (semi-)annihilation processes mentioned above, we can get the following coupled Boltzmann equations for the number densities of $\W$ and $\X$:

\begin{equation}
\begin{split}
\frac{Hz}{s}\frac{dY_\W}{dz}=&-2\sigmav(\W\W^\dagger\rightarrow\X\X^\dagger)\left(Y_\W^2-\frac{Y_\X^2}{Y_{\X,\text{eq}}^2}Y_{\W,\text{eq}}^2\right) \\
&-2\sigmav(\W\W^\dagger\rightarrow ss+\Z s)\left(Y_\W^2-Y_{\W,\text{eq}}^2\right) \\
&-\sigmav(\W\Z\rightarrow\X\X)\left(Y_\W Y_{\Z,\text{eq}}-\frac{Y_\X^2}{Y_{\X,\text{eq}}^2}Y_{\W,\text{eq}}Y_{\Z,\text{eq}}\right) \\
&-\sigmav(\W\X^\dagger\rightarrow\A\X+s\X)\left(Y_\W Y_\X-Y_\X Y_{\W,\text{eq}}\right) \\
&+\frac{1}{2}\sigmav(\X\X\rightarrow\A\W+s\W)\left(Y_\X^2-\frac{Y_\W}{Y_{\W,\text{eq}}}Y_{\X,\text{eq}}^2\right) \\
\end{split},
\end{equation}
\begin{equation}
\begin{split}
\frac{Hz}{s}\frac{dY_\X}{dz}=&2\sigmav(\W\W^\dagger\rightarrow\X\X^\dagger)\left(Y_\W^2-\frac{Y_\X^2}{Y_{\X,\text{eq}}^2}Y_{\W,\text{eq}}^2\right) \\
&-2\sigmav(\X\X^\dagger\rightarrow ss+\A s+\A\A+\Z\A+\Z s)\left(Y_\X^2-Y_{\X,\text{eq}}^2\right) \\
&+2\sigmav(\W\Z\rightarrow\X\X)\left(Y_\W Y_{\Z,\text{eq}}-\frac{Y_\X^2}{Y_{\X,\text{eq}}^2}Y_{\W,\text{eq}}Y_{\Z,\text{eq}}\right) \\
&-\sigmav(\X\X\rightarrow s\W+\A\W)\left(Y_\X^2-\frac{Y_\W}{Y_{\W,\text{eq}}} Y_{\X,\text{eq}}^2\right) \\
&+\sigmav(\Z\Z\rightarrow \X\X^\dagger)\left(Y_{\Z,\text{eq}}^2-\frac{Y_\X^2}{Y_{\X,\text{eq}}^2}Y_{\Z,\text{eq}}^2\right) \\
\end{split}.
\end{equation}

\subsection{Non-relativistic Cross sections}\label{crosssections}
We compute the tree-level non-relativistic cross section at the leading order in the relative velocity $v_{\text{rel}}$, i.e. taking into account only the $s$-wave contribution $\sigma_0$ (averaging over initial spin and gauge components, and multiplying by $\kappa=1/2$ for 2 complex DM particles in the initial state or $\kappa=1$ for other cases, as pointed out in the previous section). In this approximation, the thermal averaged cross section is simply $\sigmav=\sigma_0$.
For the group theory factors, we use the following notations:
\begin{itemize}
	\item $d_R$ is the dimension of the fundamental representation of $\Sp(\N)$, i.e. $d_R=\N$;
	\item $d_G$ is the dimension of the adjoint representation of $\Sp(\N)$, i.e. $d_G=\sfrac{\N(\N+1)}{2}$; 
	\item $C_R$ is the Casimir for the fundamental representation of $\Sp(\N)$, i.e. $C_R=\sfrac{(\N+1)}{4}$;
	\item $C_G$ is the Casimir for the adjoint representation of $\Sp(\N)$, i.e. $C_G=\sfrac{(\N+2)}{2}$; 
	\item The Dynkin Index of the fundamental representation of $\Sp(\N)$ is normalized as 1/2 as explained in section \ref{model}.
\end{itemize}
The cross sections of all the relevant processes can be expressed as:
\begin{equation}
\sigmav(\W\W^{\dagger}\to \s\s)=\frac{11M_\W^2}{288\pi w^4}=\frac{11 \g^4}{4608\pi M_\W^2},
\label{eq:1}
\end{equation}

\begin{equation}
\sigmav(\X^{\dagger}\X\to \s\s)=\frac{11M_\X^2}{288\pi d_R w^4}=\frac{11 \g^4}{4608\pi\N M_\X^2}=\frac{11 \g^4}{9216\pi\N M_\W^2},
\end{equation}

\begin{equation}
\sigmav(\W\W^{\dagger}\to \s\Z)=\frac{\g^2(4M_\W^2-M_\Z^2)^3}{576\pi M_\W^6 w^2}=\frac{3 \g^4}{256\pi M_\W^2},
\end{equation}

\begin{equation}
\sigmav(\X^{\dagger}\X\to \s\Z)=\frac{\g^2(4M_\W^2-M_\Z^2)^3}{2304\pi d_R M_\X^6 w^2}=\frac{\g^4}{1152\pi\N M_\W^2},
\end{equation}

\begin{equation} \label{xxglueballs}
\sigmav(\X^{\dagger}\X\to \A\A)=\frac{19C_R(4C_R-C_G)\g^4}{576\pi d_R M_\X^2}=\frac{19 \g^4(\N+1)}{2304\pi M_\W^2},
\end{equation}

\begin{equation}
\sigmav(\X^{\dagger}\X\to \A\s)=\frac{C_R\g^2}{9\pi d_R w^2}=\frac{\g^4(\N+1)}{144\pi\N M_\W^2},
\end{equation}

\begin{equation}
\sigmav(\W\W^{\dagger}\to \X^{\dagger}\X)=\frac{225 \g^4 d_R}{8192\sqrt{2}\pi M_\X^2}=\frac{225 \g^4\N}{4096\sqrt{2}\pi M_\W^2},
\end{equation}

\begin{equation}
\sigmav(\W\Z\to \X\X)=\frac{5 \g^4 d_R}{576\sqrt{2}\pi M_\X^2}=\frac{5 \g^4\N}{288\sqrt{2}\pi M_\W^2},
\end{equation}

\begin{equation}
\sigmav(\X\X\to \s\W)=\frac{\g^2(4M_\X^2-M_\W^2)^3}{1152\pi d_R M_\X^6 w^2}=\frac{\g^4}{576\pi\N M_\W^2},
\end{equation}

\begin{equation}
\sigmav(\W\X^{\dagger}\to \s\X)=\frac{(30-13\sqrt{2}) \g^4}{1152\pi M_\W^2},
\end{equation}

\begin{equation}
\sigmav(\Z\Z\to \X^{\dagger}\X)=\frac{1705 \g^4 d_R}{18432\sqrt{2}\pi M_\Z^2}=\frac{1705 \g^4\N}{18432\sqrt{2}\pi M_\W^2},
\end{equation}

\begin{equation}
\begin{split}
\sigmav(\X^{\dagger}\X\to \Z\A)&=\frac{g^4 d_G }{576\pi M_\X^2 d_R^2}\left(1-\frac{M_Z^2}{4M_\X^2}\right)\left(19+\frac{M_\Z^2}{M_\X^2}+\frac{M_\Z^4}{4M_\X^4}\right) \\
&=\frac{11 C_R \g^2}{36\pi d_R w^2}=\frac{11 \g^4(\N+1)}{576\pi\N M_\W^2}
\end{split},
\end{equation}

\begin{equation}
\sigmav(\X\X\to \A\W)=\frac{11 C_R \g^2}{18\pi d_R w^2}=\frac{11 \g^4(\N+1)}{288\pi\N M_\W^2},
\end{equation}

\begin{equation}
\sigmav(\W\X^{\dagger}\to \A\X)=\frac{(32-21\sqrt{2}) C_R \g^4}{72\pi  M_\W^2}=\frac{(32-21\sqrt{2}) \g^4(\N+1)}{144\pi M_\W^2}.
\label{eq:11}
\end{equation}	
As a check of our computations, we have verified that all the cross sections scale as 1/$s$ in the ultra-relativistic limit\footnote{For $\N=0$, cross sections involving $\A$ and $\X$ vanish, $\Z$ becomes DM forming a degenerate triplet with $\W$, and the results in~\cite{Hambye:2008bq } is reproduced taking into account the extra cross sections $2\sigmav(\W\W^{\dagger}\to\s\s)=\sigmav(\Z\Z\to\s\s)$ and $2\sigmav(\W\W^{\dagger}\to\Z\s)=\sigmav(\W\Z\to\W\s)=3\g^4/128\pi M_\W^2$.}.

\subsection{Relic abundance}\label{relicsection}
\label{relic}

If the dynamics is purely perturbative, the DM abundance is given by the sum of $\W$ and $\X$ abundances, and one can use the following expressions (the subscript 0 means quantities evaluated at the present time):
\begin{equation}
\Omega_{\text{DM}}=\frac{\rho_{\text{DM}}}{\rho_{\text{cr}}}=\frac{s_0}{3H_0^2/8\pi G}\left(M_\W Y_{\W0}+M_\X Y_{\X0}\right)
\end{equation}
or
\begin{equation}
\frac{\Omega_{\text{DM}}h^2}{0.110}=\frac{M_\W}{0.4  \text{eV}}\left(Y_{\W0}+\frac{Y_{\X0}}{\sqrt{2}}\right).
\end{equation}
Instead, in our case the Sp($\N$) dynamics confines at the scale $\Lambda_\text{DC}$, given by eq.\eqref{lambdaDC}.
After the confinement, $\X$ can form two kinds of dark mesons: (1) the meson $\X^\dagger\X$ (and $\X^\dagger \D_{\mu} \X$), which carry no dark-baryon number and decay into glueballs through the $\X\X^{\dagger}\A\A$ vertex; (2) the mesons $\mathcal{M}\sim \X^T\gamma_\N\mathcal{D}_\mu\X$ with dark-baryon number 2. The cubic vertex $\X\X\W^{\dagger}$ becomes a $\mathcal{M}\W$ mass mixing. In the limit of $\Lambda_{\text{DC}}\ll w$, their masses are $M_{\mathcal{M}}\simeq\sqrt{2}M_{\W}$ so that $\mathcal{M}$ will 
decay into $\W$ and glue-balls through the $\A\X\X\W^{\dagger}$ vertex (we assume that decays are faster than $\mathcal{M}^*\mathcal{M}$ annihilations), leaving $\W$ as the only DM candidate.
\newline

We take into account the confinement of $\Sp(\N)$ dynamics and the corresponding hadronization under the following simplified assumptions:
\begin{itemize}
	\item confinement of $\Sp(\N)$ dynamics occurs after the perturbative freeze-out of $\W$ and $\X$, i.e. $\Lambda_\text{DC}\ll T_f\sim M_\W/25$ (this is roughly the correct order of magnitude for $T_f$, and can be verified through the numerical solutions of Boltzmann equations as shown in figure\fig{DMabundancefig}). This gives an upper limit on $\g$ that is, using eq.\eqref{lambdaDC}:
	\beq \g^2\ll\frac{48\pi^2}{11(\N+2)\log25}. \eeq 
	In the opposite regime confinement occurs before DM decoupling and invalidates our computations. 
	\item about half $\X$ form charged mesons $\mathcal{M}$: whenever two  $\X$ meet, they can form either the meson $\X^{\dagger}\X$ or the meson $\mathcal{M}\sim\X\X$, and each case has a 50\% chance. However, only  mesons $\mathcal{M}$ contribute to $\W$ abundance.
	\item all the mesons $\mathcal{M}$ decay to $\W$ (+ glueballs/scalon): these processes occurs through $\X\X\A\W$ and other vertices. As we have already pointed out, we assume that  $\mathcal{M}$ decays always faster than  $\mathcal{M}^*\mathcal{M}$ annihilations.  
\end{itemize}
Under these assumptions, the DM relic abundance is given by
\begin{equation}\label{DMabundance}
\frac{\Omega_{\text{DM}}h^2}{0.110}=\frac{M_\W}{0.4  \text{eV}}\left(Y_{\W0}+\frac{Y_{\X0}}{4}\right).
\end{equation}

\subsubsection*{Sommerfeld enhancement and bound states}

In this subsection we dicuss the effects of the Sommerfeld enhancement factors and (perturbative) bound states formation on the relic abundance. Following the analysis in section \ref{Sommerfeld}, we compute the Sommerfeld enhancement factors for the (semi-)annihilation cross sections in \ref{crosssections}. 
Then we solve the Boltzmann equations to get the DM relic abundance as explained in section \ref{relicsection} and we compare the results obtained (i) with the enhancement factors and (ii) without the enhancement factors.  We find that the difference between (i) and (ii) is neglible ($\lesssim3\%$). So we conclude that Sommerfeld enhancement gives only small corrections to the relic abundance and can be safely ignored (notice that the contribution of the Sommerfeld enhancement to indirect detection cross section is instead not neglibile as we explain in section \ref{indirectsection}).

In principle, the formation of Coulombian bound states from $\X\X^{\dagger}$($\X\X$) states through the exchange of dark gluons $\A$ (discussed in section \ref{Sommerfeld}) can also affect the relic abundance. However, the order of magnitude of this effect is roughly comparable to Sommerfeld corrections. Generally speaking, the bound states formation is relevant whenever Sommerfeld enhancement is significant. As we discussed above, this is not the case for our model. Therefore, we can reasonably neglect these Coulombian bound states in our computation.

As a final remark we notice that the effect of both the Sommerfeld enhancement factors and the perturbative bound state formation on the relic abundance is reasonably subleading if compared to the uncertanties due to the process of hadronization that occours at $T\simeq\Lambda_{\text{DC}}$, when the theory confines, which we discussed in section \ref{relic}. 

\subsubsection*{Glueballs dilution}
Dark glueballs (DG) can decay to SM particles through the Higgs portal. The one-loop effective interaction between the scalon $\s$ and the dark gluons $\A$ has been computed in \cite{Buttazzo:2019iwr} as:
\begin{equation}\label{Leff}
\mathcal{L}_{\text{eff}}^{\s\A\A}=-\frac{7\alpha}{16\pi^2}(\A_{\mu\nu})^2\left(\frac{\s}{w}-\frac{\s^2}{2w^2}+...\right)
\end{equation}
with $\alpha=\g^2/4\pi$.
The lifetime of glueballs can be estimated as $\tau_{\text{DG}}=\GammaDG^{-1}$, which depends on the mass ordering. If the mass of the glueballs is larger than the weak scale, they decay into Higgs components as \cite{Buttazzo:2019iwr}:
\begin{equation}
\GammaDG(\text{DG}\to\s\to H^{\dagger}H=hh+ZZ+WW)=\frac{49 f_{\text{DG}}^2\alpha^2 \lambda_{H\S}^2}{2048\pi^3 M_{\text{DG}}M_\s^4}\text{Re}\sqrt{1-\frac{4M_{h,W,Z}^2}{M_{\text{DG}}^2}}.
\end{equation}
The glueballs life-time can instead become cosmologically large if glueballs are enough lighter than the weak scale; in this case we get \cite{Buttazzo:2019iwr}:
\begin{equation}\label{glueballdecay}
\GammaDG=\left(\frac{7\alpha f_{\text{DG}}}{32\pi w}\right)^2\sin^22\gamma\left(\frac{1}{M_{S_1}^2-M_{\text{DG}}^2}-\frac{1}{M_{S_2}^2-M_{\text{DG}}^2}\right)^2\Gamma_{h_{\text{DG}}},
\end{equation}
where $M_{\text{DG}}\approx 7\Lambda_{\text{DC}}$ is the mass of the glueball~\cite{Morningstar:1999rf}, $f_{\text{DG}}\sim M_{\text{DG}}^3$ is a dark matrix element, $\Gamma_{h_{\text{DG}}}$ is the decay width of a SM Higgs with mass $M_{\text{DG}}$, $\gamma$ is the mixing angle that diagonalises the Higgs-scalon mass matrix, $M_{S_i}$ are the mass eigenvalues. In more detail we have:
\begin{equation}\label{sin2gamma}
\sin2\gamma=\frac{v^2\sqrt{8\lambda_H\lambda_{H\S}}}{{M^2_{S_2}}-{M^2_{S_1}}}
\end{equation}
with ${M_{S_2}}\approx M_{\s}=\sqrt{\beta_{\lambda_\S}}w$ and ${M_{S_1}}\approx M_h\simeq125$ GeV.
If the lifetime is long enough ($\tauDG\gg t_{\Lambda_{\text{DC}}}$, where $t_{\Lambda_{\text{DC}}}$ is the time at which confinement occurs, estimated by $H(T=\Lambda_{\text{DC}})=t_{\Lambda_{\text{DC}}}^{-1}$), glueballs can dominate the energy density of the Universe while decaying into SM particles, and the reheating effect dilutes the DM density as $Y_{\text{diluted}}=YD$~\cite{Mitridate:2017oky}. The dilution factor is estimated as:
\begin{equation}\label{dilution}
D^{-1}=\left[1+\frac{g_{\text{DG}}}{g_{\text{SM}}^{2/3}}\left(\frac{{\Lambda_{\text{DC}}^2}}{\GammaDG M_{P}}\right)^\frac{2}{3}\right]^{3/4}
\end{equation}
with $g_{\text{SM}}=106.75$ and $g_{\text{DG}}=\N(\N+1)$. This effect is taken into account into the computation by multiplying the r.h.s. of eq.(\ref{DMabundance}) by a factor of $D$.
However, it becomes irrelevant if the lifetime is very short.

We define the mixing angle $\epsilon$ between the Higgs and the dark glueball from eq.(\ref{glueballdecay}):
\begin{equation}
\epsilon=\frac{7\alpha f_{\text{DG}}}{32\pi w}\sin2\gamma\left(\frac{1}{M_{S_1}^2-M_{\text{DG}}^2}-\frac{1}{M_{S_2}^2-M_{\text{DG}}^2}\right).
\label{eq:mix}
\end{equation}
In figure\fig{Scalonfig} we show the predicted Higgs-glueball mixing angle $\epsilon^2$ as a function of the glueball mass $M_{\text{DG}}$ with an assumption that some extra new physics gives fast glueball decays through different channels. We also show the current limits at LHC (shaded regions) and the future sensitivity of SHiP experimental proposal\cite{Alekhin:2015byh} (dashed line).

\begin{figure}[!t]
	\label{DMabundancefig}
	\centering
	$$\includegraphics[width=0.42\textwidth]{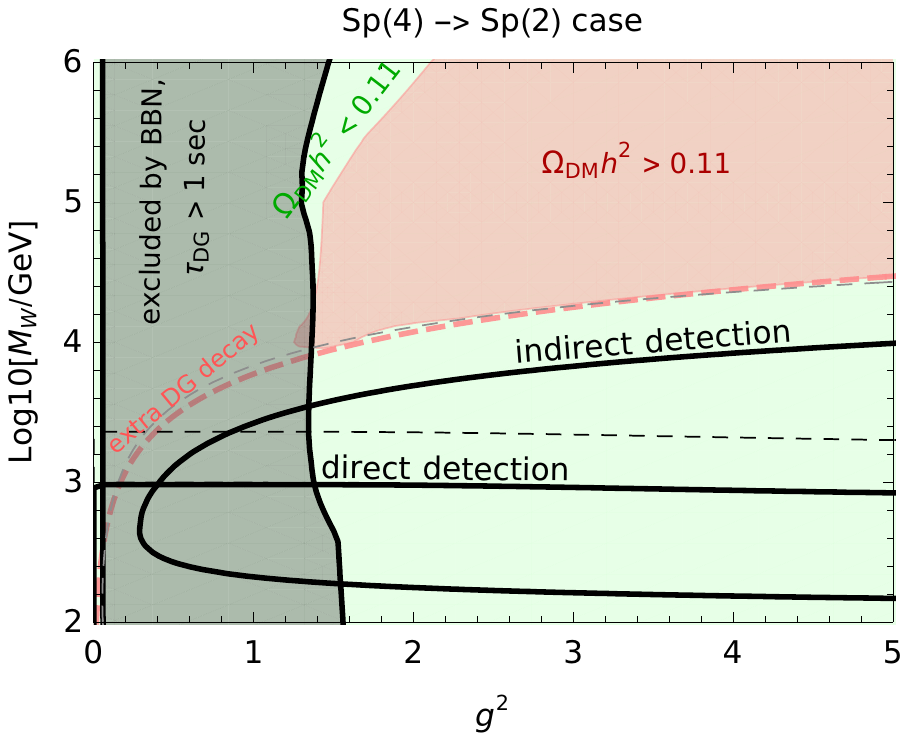}\qquad~~~~~
	\includegraphics[width=0.42\textwidth]{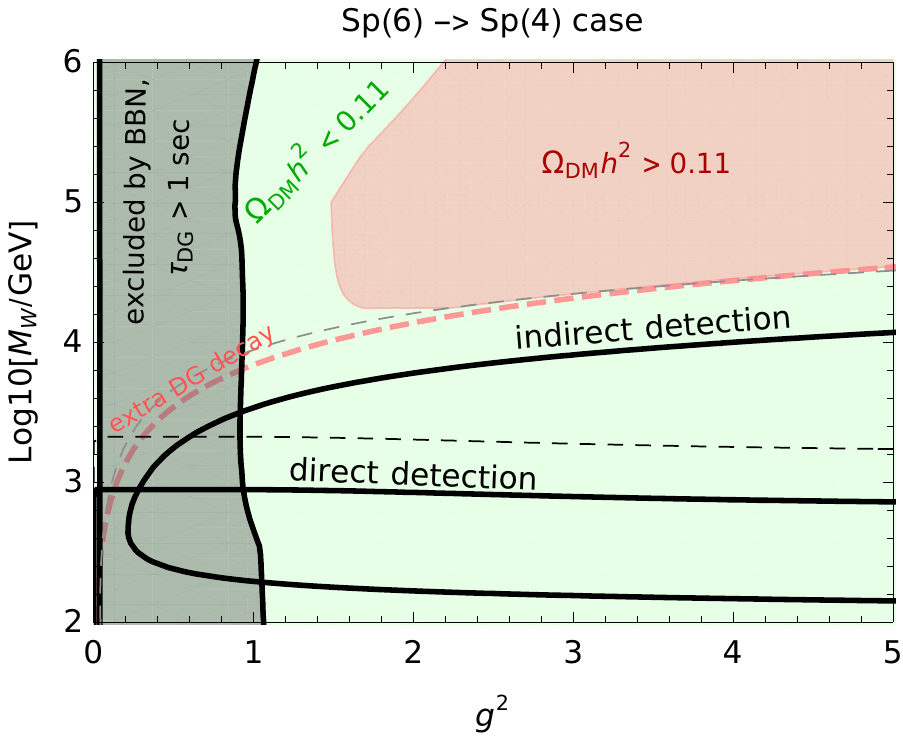}$$\\
	\vspace{-1cm}
	$$\includegraphics[width=0.42\textwidth]{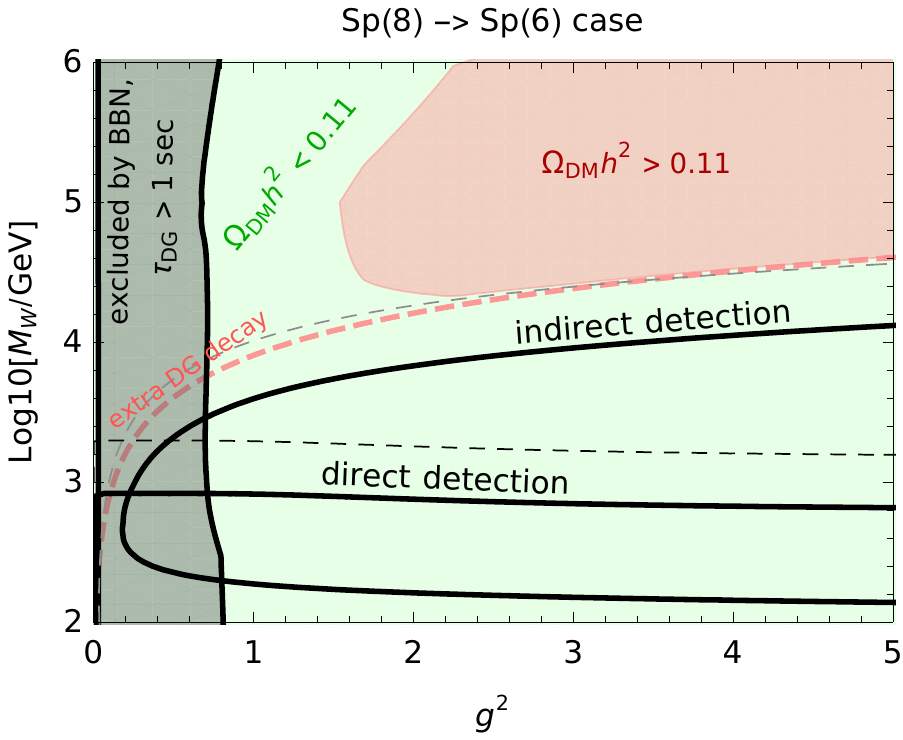}\qquad
	\includegraphics[width=0.466\textwidth]{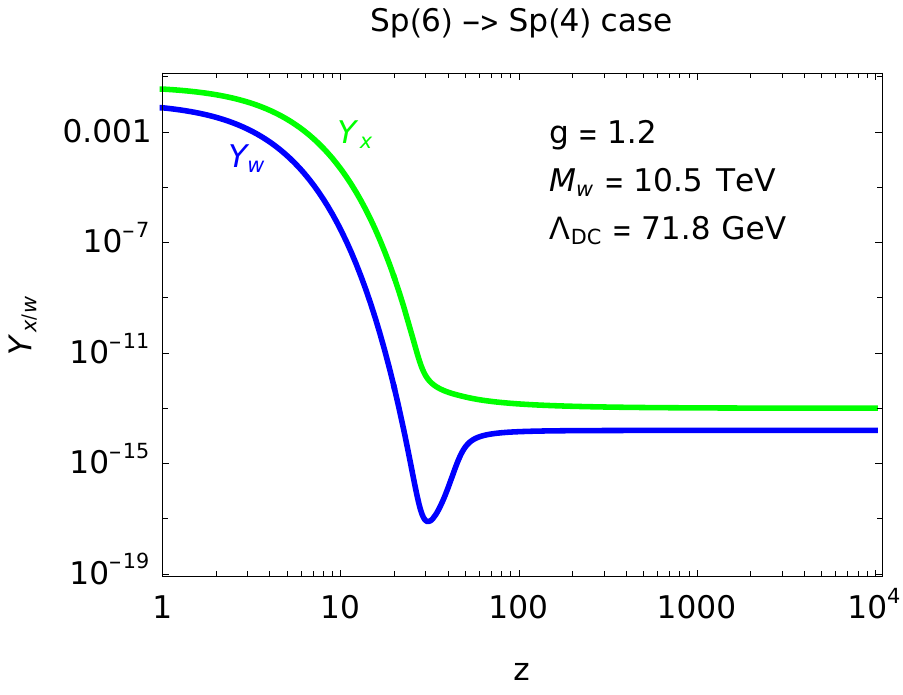}$$
	\vspace{-1cm}
	\caption{\em \label{fig:DMabundancefig} 
		The DM abundance is reproduced ($\Omega_{\text{DM}}h^2=0.11$) along the boundaries between the red/green regions: the red-solid line is obtained taking into account the dilution factor due to the long-lived glueballs, while the red-dashed line is computed with the assumption that some extra new physics gives fast glueball decays. The current (future) limits of direct and indirect detections are also shown in black solid (dashed) line, respectively. The left gray region is excluded by BBN because of too slow DG decays (of course it is not excluded if we assume extra DG decays). We show predictions for different values of $\N$. In the last plot we show the solutions of the Boltzmann equations for values of the parameters ($\N=4$, $\g = $1.2, $M_{\W} = $10.5 TeV) such that the cosmological DM abundance is reproduced. We find that the freeze-out temperature is roughly $T_f\sim M_{\W}/25\gg\Lambda_{\text{DC}}$, which is in agreement with our assumptions.	}
\end{figure}

The predicted DM relic density as a function of $\g^2$ and $M_{\W}$ is plotted in figure\fig{DMabundancefig}. In the red region the overall DM abundance turns out to be larger than the cosmological value ($\Omega_{\text{DM}}h^2=0.11$), while the green region indicates $\Omega_{\text{DM}}h^2 < 0.11$.
The red-solid line is obtained by 
numerical integration of the Boltzmann equations and taking into account the dilution factor in eq.~(\ref{dilution}) due to long-lived glueballs.
We also show, as the red dashed line, the result obtained by assuming that some extra new physics gives fast glueball decays (i.e. no dilution factor).
Besides, the limitations of current (future) direct~\cite{Aprile:2018dbl} and indirect~\cite{Ackermann:2015zua,Abdallah:2016ygi} detections are also displayed with black solid (dashed) lines, respectively.
Future perspective on indirect detection (CTA)~\cite{Morselli:2017ree} can explore a significant region of the parameter space
 especially in the case in which glueballs decay very fast.

In figure\fig{DMabundancefig} (bottom right corner) we show the solutions of the Boltzmann equations for values of the parameters such that the DM abundance is reproduced. We see some modifications with respect to the conventional WIMPs behaviour. In particular at freeze-out (roughly $z\sim25$) the $\W$ abundance is initially highly suppressed with respect to $\X$ because of the mass gap between the two states, but it suddenly increases because of the conversion processes $\X\X\to\s\W$ and $\X\X\to\A\W$. 

\section{Signals and bounds}\label{pheno1}
\subsection{Scalon production}
\begin{figure}\label{scalonfig}
	\centering
	$$\includegraphics[width=0.40\textwidth]{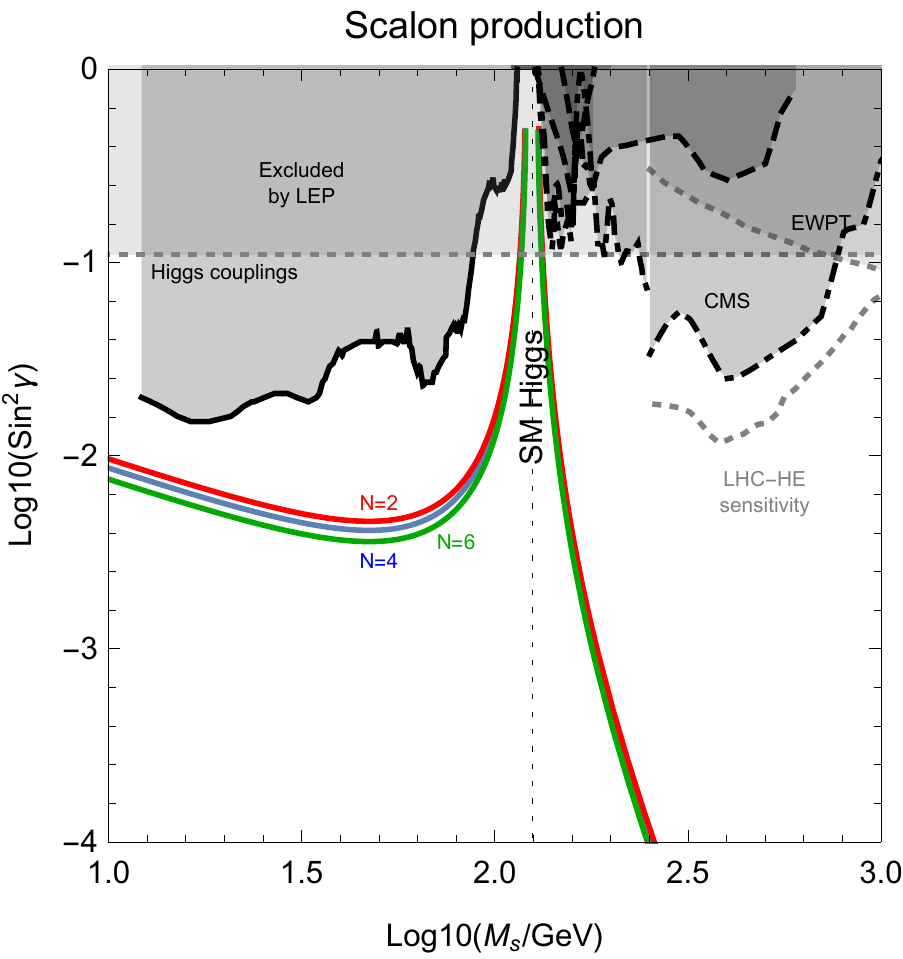}\qquad
	\includegraphics[width=0.46\textwidth,height=0.416\textwidth]{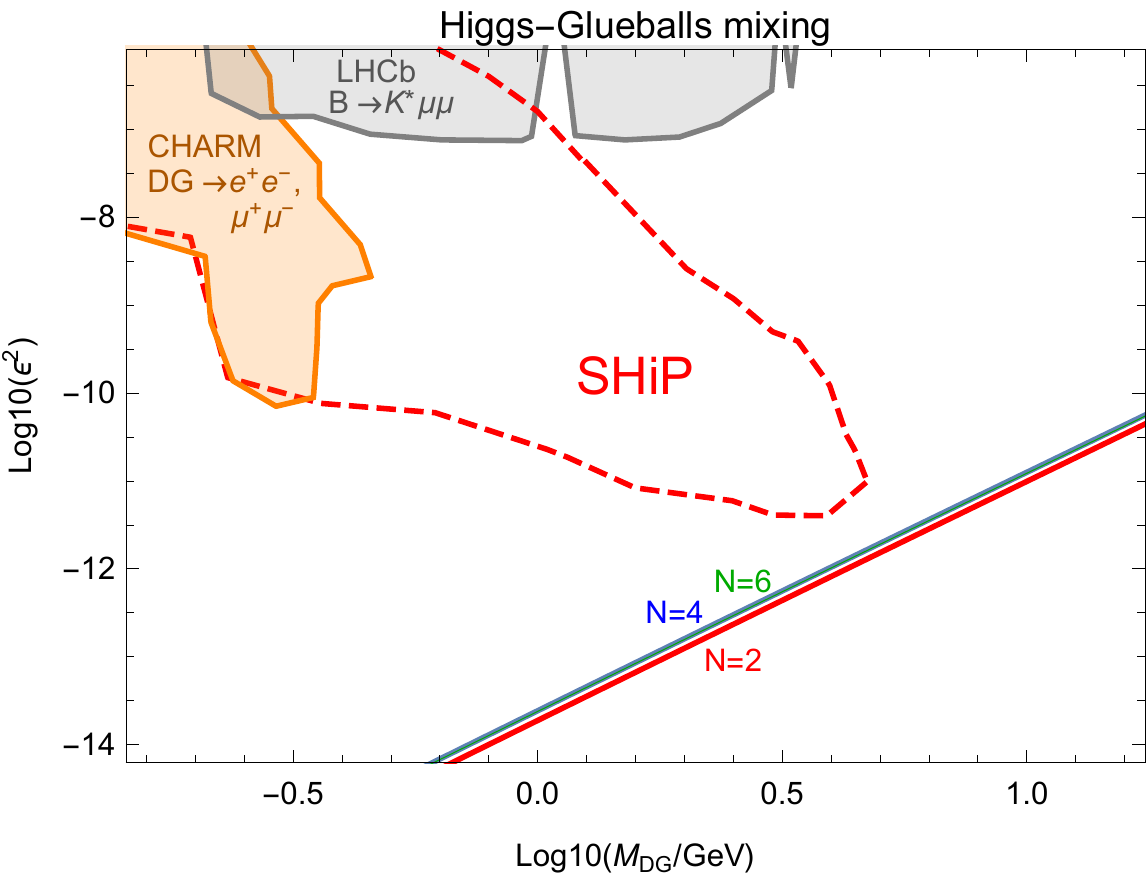}
	$$
	\vspace{-1cm}
	\caption{\em \label{fig:Scalonfig} 
		On the left: predicted production cross section for the scalon, under the assumption that $\W$ reproduce the full DM abundance of the Universe and some extra new physics gives fast glueball decays. Signals are very supressed if glueballs decay slowly. On the right: predicted Higgs-glueball mixing angle as a function of the glueball mass $M_{\text{DG}}$ with the same assumptions.}
\end{figure}

The scalon $\s$ behaves like an extra Higgs boson with couplings rescaled by $\sin\gamma$, where the mixing angle $\gamma$ is given by eq.\eqref{sin2gamma}. We show in figure\fig{Scalonfig} (left) the predicted value of $\sin^2\gamma$, which is equal to the production cross section for the scalon in SM Higgs cross section units, as a function of the scalon mass $M_\s$. We assume that some extra new physics gives fast glueball decays (if glueballs decay slowly the production cross section is supressed by many order of magnitude and not testable at current/future experiments).
Various present and projected constraints from Higgs measurements and direct searches are also shown \cite{Buttazzo:2015bka,Buttazzo:2018qqp}. In particular, it can be seen that measuring Higgs
couplings with a $10^{-3}$ precision, which can be attained at future lepton colliders, would allow
to probe a significant region of the parameter space.
\subsection{Direct detection}
\begin{figure}\label{DDfig}
	\centering
	$$\includegraphics[width=0.40\textwidth]{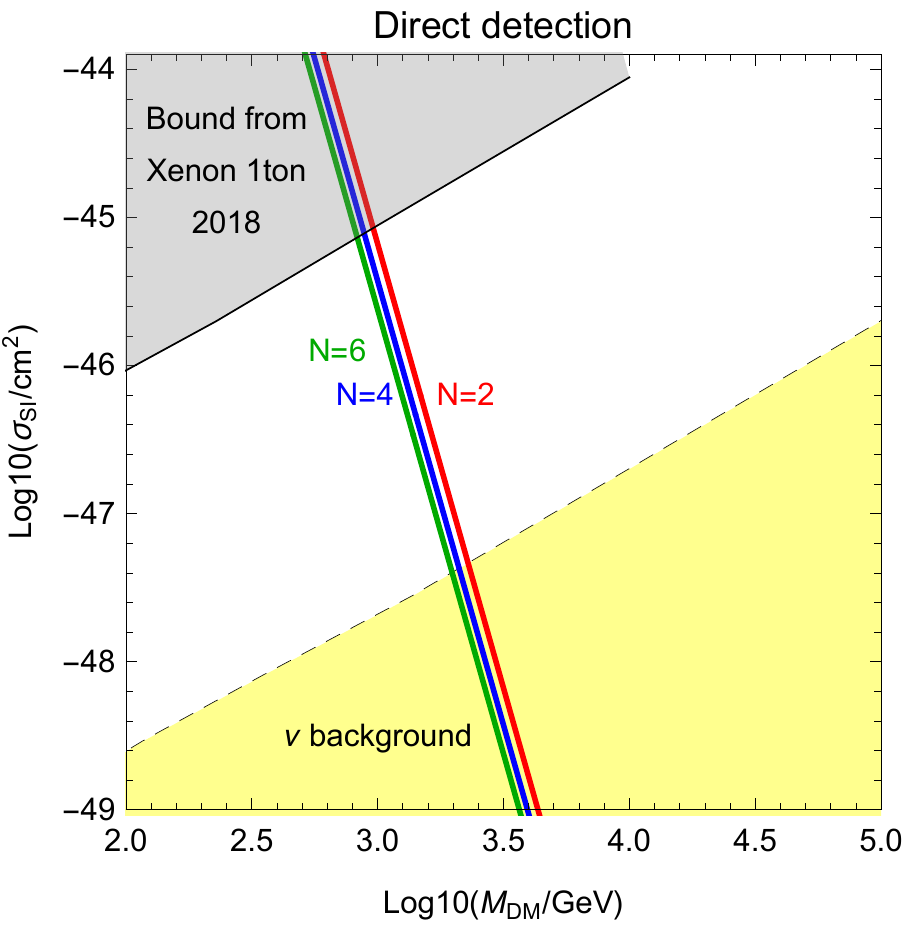}\qquad
	\includegraphics[width=0.40\textwidth]{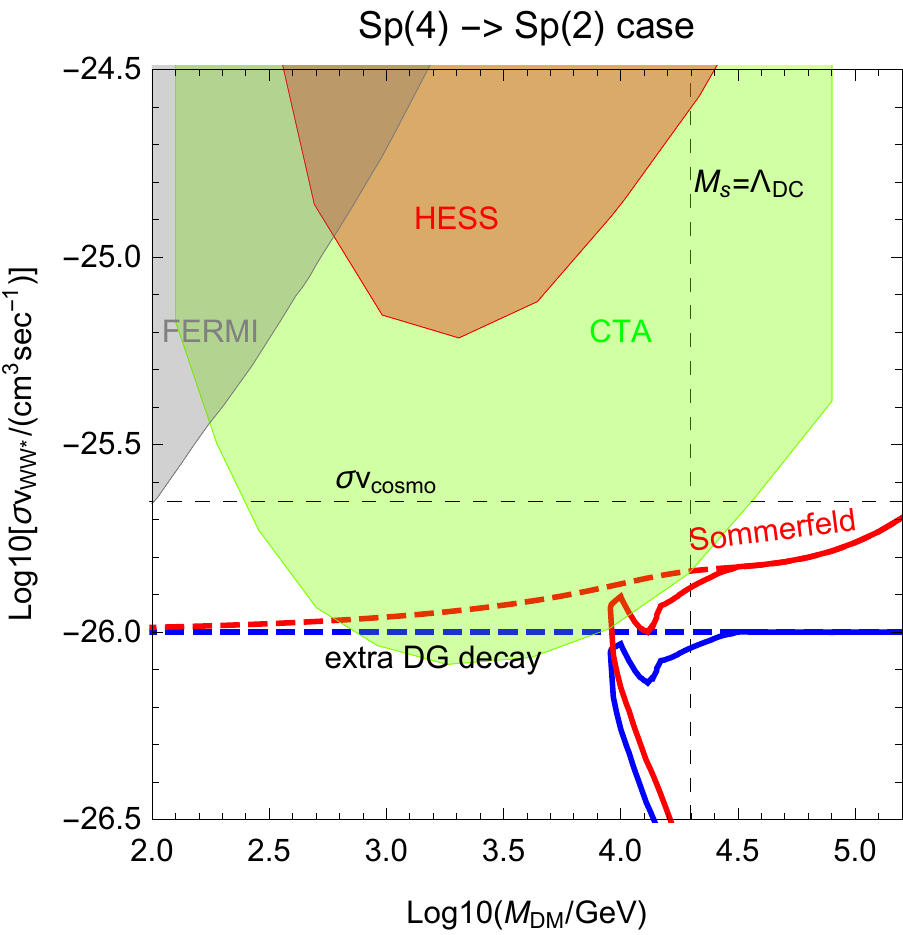}
	$$
	\vspace{-1.cm}\\
	$$\includegraphics[width=0.40\textwidth]{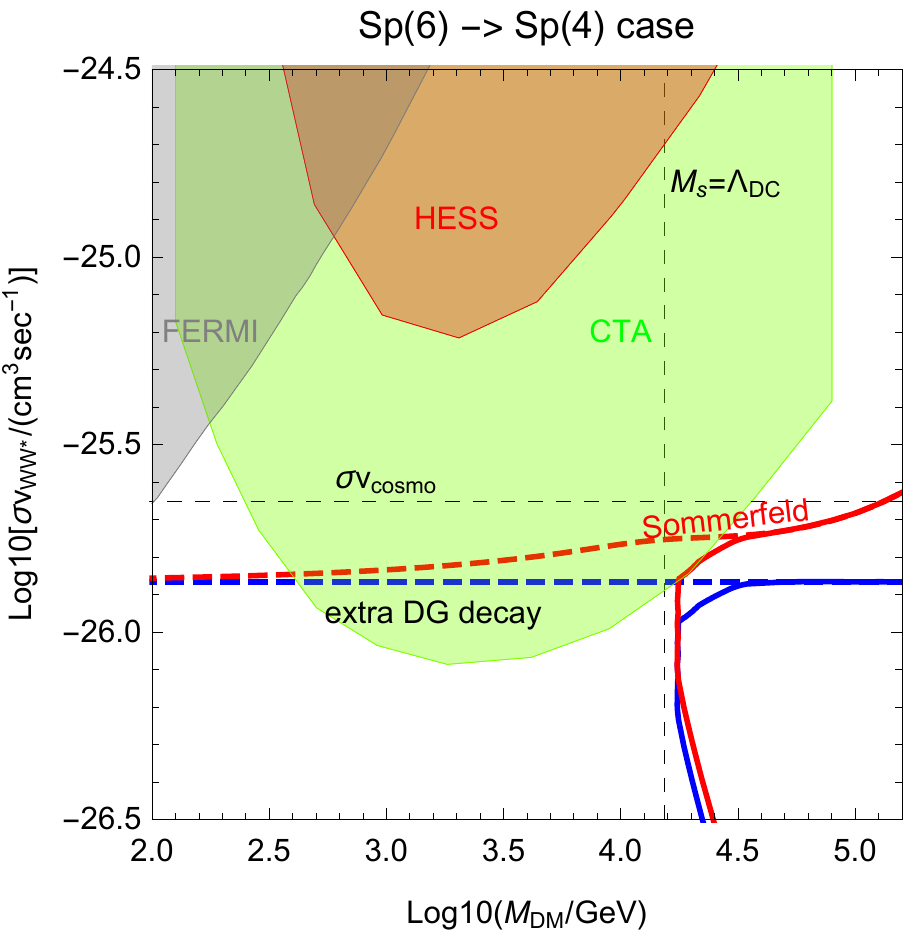}\qquad~
	\includegraphics[width=0.40\textwidth]{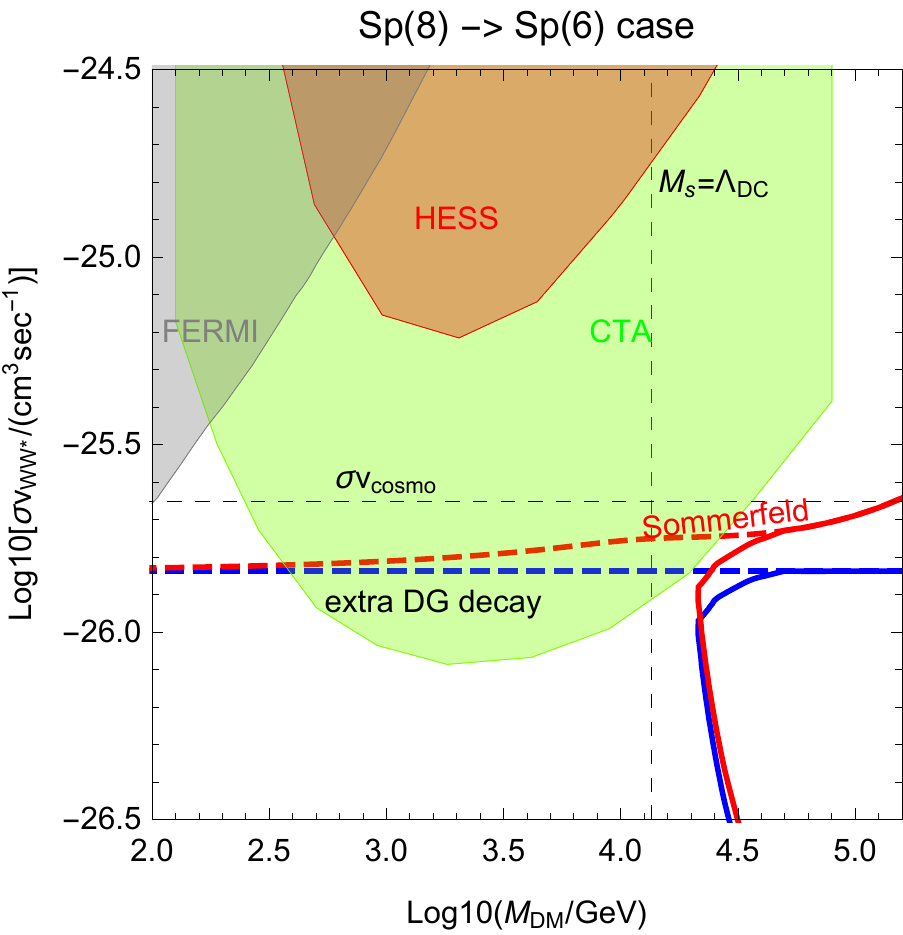}~~$$
	\vspace{-1cm}
	\caption{\em \label{fig:DDfig} 
		The first plot: predicted cross section for DM direct detection under the assumption that $\W$ reproduce the full DM abundance of the Universe and some extra new physics gives fast glueball decays. Other plots: predicted annihilations cross sections for indirect detection under the assumption that $\W$ reproduce the full DM abundance of the Universe, for different values of $\N$. The solid lines are obtained taking into account the dilution due to long-lived glueballs; the dashed lines are obtained assuming that some extra new physics gives fast glueball decays. The red lines take into account the Sommerfeld enhancement in indirect detection cross section. The vertical dashed line is the boundary between the perturbative and condensed phase, which is estimated as the scale at which $M_\s\approx \Lambda_{\text{DC}}$.}
\end{figure}

As pointed out in \cite{Buttazzo:2019iwr} the spin-indipendent cross section for direct detection is the same as in the $\SU(2)$ model, and it can be expressed as
\beq
\sigma_{\text{SI}}=\g^2\sin^22\gamma\frac{m_N^4f^2}{16\pi v^2}\left(\frac{1}{M^2_{S_1}}-\frac{1}{M^2_{S_2}}\right)^2,
\label{eq:direct}
\eeq
where $f\simeq0.3$ is a nuclear matrix element.
In figure\fig{DDfig} (top left corner) we plot the DM direct detection cross section as a function of the DM mass under the assumptions that $\W$ reproduce the full DM abundance and some extra new physics gives fast glueball decays. We also show the constraints from Xenon1T~\cite{Aprile:2018dbl} as well as neutrino floor.
If glueballs decay very slowly, the direct detection cross section is supressed and not observable at current and future experiments, as we can observe in figure\fig{DMabundancefig}. 

\subsection{Indirect detection}\label{indirectsection}
After the confinement of the gauge subgroup $\Sp(\N)$ below the energy scale $\Lambda_{\text{DC}}$, $\W$ becomes the only DM candidate.
When $\W$ and $\W^{\dagger}$ annihilate with each other, they can produce indirect detection signals, which can be expressed as
\begin{equation}
\sigma_{\W\W^{\dagger}}v_{\text{rel}}=
\sigmav(\W\W^{\dagger}\to\s\s, \s\Z, \text{dark hadrons}),
\end{equation}
where the dark hadrons are glueballs ($\A\A$ bound state) and the uncharged mesons ($\X^{\dagger}\X$ bound state). The energy of the scattering processes is roughly
$M_\W\gg\Lambda_{\text{DC}}$, so we can use the perturbative results in section~\ref{crosssections}. In particular $\sigmav(\W\W^{\dagger}\to\A\A)$ is one-loop supressed, so the relevant cross section into dark hadrons is given by $\sigmav(\W\W^{\dagger}\to \X^{\dagger}\X)$. 
Annihilations products will further decay into SM particles. The total annihilation cross section relevant for indirect detection is:
\begin{equation}
\sigma_{\W\W^{\dagger}}v_{\text{rel}}=\frac{\g^4}{256\pi M_\W^2}\left(\frac{65}{18}+\frac{225}{16\sqrt{2}}\N\right).
\end{equation}
We take into account the Sommerfeld enhancement considering the scalon exchange among $\W$ and $\W^{\dagger}$ through the $\s\W\W^\dagger$ interactions. 
The cross section get enhanced as
\begin{equation}
(\sigma_{\W\W^{\dagger}}v_\text{rel})_{\text{enhanced}}=S\sigma_{\W\W^{\dagger}}v_\text{rel}.
\end{equation}
For $v_\text{rel}\ll M_\s/M_\W$, which is the relevant regime for indirect  detection, the eq.\eqref{eq:sommer1} can be further simplified as~\cite{Mitridate:2017izz}
\begin{equation}
S=\frac{2\pi^2\alpha_\text{eff} M_\W}{\kappa M_\s}\left(1-\cos2\pi\sqrt{\frac{\alpha_\text{eff} M_\W}{\kappa M_\s}}\right)^{-1}
\end{equation}
with $\kappa\simeq1.74$ and $\alpha_\text{eff}=\g^2/16\pi$.
The indirect detection cross section is not affected by formation of perturbative bound states: indeed, as we have pointed out at the end of section \ref{Sommerfeld}, the potential \eqref{potential} doesn't allow for the existence of $\W\W^{\dagger}$ bound states, for the values of the parameters of the model.

In figure\fig{DDfig} we plot the DM annihilation cross section as a function of the DM mass under the assumption that $\W$ reproduce the full DM abundance. The solid lines are obtained by taking into account the dilution due to long-lived glueballs, while the dashed lines assume that some extra new physics gives fast glueball decays. The blue lines have been obtained without taking into account the Sommerfeld enhancement, which is instead included in the red lines. The vertical dashed line is the boundary between the perturbative and condensed phase, which is estimated as the scale at which $M_\s\approx \Lambda_{\text{DC}}$.
We compare the annihilation cross section with the HESS and FERMI-LAT limits on gamma-ray signals~\cite{Ackermann:2015zua,Abdallah:2016ygi}. The resulting bounds are plotted in figures\fig{DMabundancefig} and\fig{DDfig}. We also show future prospects at CTA~\cite{Morselli:2017ree}. In the region excluded by indirect detection the predicted DM abundance is much smaller than the cosmological abundance. However, the CTA prospects can test a significant region of the parameter space, especially in the case in which glueball decays are very fast.

We notice that the indirect detection cross section turns out to be slightly suppressed with respect to the standard WIMPs scenario (in which one gets approximately the value $\sigmav_\text{cosmo}\simeq2.2\times 10^{-26}\text{cm}^3/\text{sec}$). 
The suppression factor depends on $\N$: it is roughly a factor of 2 for small values of $\N$ (such as the ones plotted in fig.\fig{DDfig}), while it becomes negligible for larger values of $\N$.
This is a characteristic prediction of our model, which is testable by future experiments at CTA. The suppression factor becomes smaller if one includes the Sommerfeld enhancement (red lines). We notice that if we take the glueballs dilution effects into consideration (solid lines), the indirect detection cross section is further supressed by one or more order of magnitudes in a certain range of the parameters space. 

\section{A few comments on the scale invariant hypothesis}\label{sec:scale_inv}

In the previous sections we have computed the mass spectrum and studied the phenomenology of the model within the scale invariant hypothesis $M_\S^2=0$.
Under this assumption the SSB is driven by the one-loop effective potential (Coleman-Weinberg mechanism). Furthermore, we have also dropped the Higgs mass term by demanding that the weak scale $v$ is generated by the dynamical
scale $w$ (see eq.\eqref{vev}). 
Now we would like to briefly discuss a more generic case by abandoning the scale invariant framework, and investigate its effects on the phenomenological
results. 

If $M_\S^2$ is sufficiently small (positive or negative) the dynamics is spontaneously broken by Coleman-Weinberg mechanism as in the scale invariant case. Otherwise, if the quadratic term $M_\S^2$ is not negligibly small the picture can change significantly. We distinguish two cases: (1) the quadratic term is negative $M_\S^2<0$; (2) the quadratic term is positive $M_\S^2>0$.

In the first case, if $M_\S^2$ is not too small, the $\Sp(\N+2)$ dynamics is no longer spontaneously broken and confines at low energy ($\Lambda_{\text{DC}}$). The mass spectrum in the confined phase is composed by unstable glueballs and $\S$ bound states (some of which are stable) as discussed in \cite{Buttazzo:2019iwr}. As we have pointed out in section \ref{intro} there are strong indications of a duality between this phase and the Higgsed one.

In the second case the $\Sp(\N+2)$ dynamics is already spontaneously broken at tree level. The mass spectrum is the same as we discussed in section \ref{modelR} with an important difference: the scalon mass is no longer one-loop suppressed but gets a tree-level value $M_\s^2=2\lambda_\S w^2$ (assuming a small $\lambda_{H\S}$). Therefore, if $\lambda_\S$ is not too small (e.g. $\lambda_\S>\g^4/(4\pi)^2$) the scalon can be significantly heavier than in the scale  invariant case, and this can affect the phenomenological results of the model.

Concerning the cosmological evolution, the presence of a heavier scalon can reduce the cross sections that contain the scalon in the final state due to the phase space suppression. However, the contributions of the processes that contain no scalon are dominant (see eq.\eqref{eq:1}$\sim$\eqref{eq:11}), so the mass term has very small effects ($\sim\mathcal{O}(1\%)$) on the relic density.


Both the direct detection cross section $\sigma_{\text{SI}}$ and the Higgs-glueballs mixing $\epsilon^2$ get suppressed by a heavier scalon (see eq.\eqref{eq:mix}$\sim$\eqref{eq:direct}). 
As for the indirect detection, the processes $\W\W^\dagger\to\s\s,\s\Z$ are suppressed (even kinematically forbidden if the scalon is very heavy), while the process $\W\W^\dagger\to\X\X^{\dagger}$ (predominant) is unaffected, so the indirect detection cross section is roughly the same as in the scale invariant case.

Finally we notice that, if we drop the scale invariant hypothesis, the model contains 4 free parameters ($M_S^2,~\lambda_{H\S},~\g,~\lambda_\S$). On the contrary, for the scale invariant case there are only 2 independent parameters left, so the phenomenological analysis is less generic but more predictive.

\section{Conclusions}\label{conclusions}
Motivated by the previous work~\cite{Buttazzo:2019iwr}, we have studied the phenomenology of Sp DM model, which contains a new dark gauge group $\Sp(\N+2)$ that is spontaneously broken to $\Sp(\N)$ by a scalar $\S$ in the fundamental representation. We further restricted the parameter space assuming that:
\begin{itemize}
	\item the cosmological DM abundance is reproduced thermally;
	\item the  $\Sp(\N+2)\to \Sp(\N)$ symmetry breaking occurs dynamically à la Coleman-Weinberg;
	\item the $\S$ vacuum expectation value also induces the observed Higgs mass.
\end{itemize} 
Thanks to these extra assumptions, DM phenomenology is described by only one free parameter, the DM mass $M_\W$ (or  the dark gauge coupling $\g$). We have taken into account the confinement of $\Sp(\N)$ dynamics and the corresponding formation of bound states.  Of course, some of the above assumptions can be relaxed, giving more general phenomenology.

In section \ref{modelR} we have summarized the main features of the model, including the Lagrangian before and after the symmetry breaking, the perturbative mass spectrum and the formation of bound states.
The theory admits an accidental global dark-baryon number $\U(1)$ symmetry, which renders two co-stable vector DM candidates at the perturbative level: $\W$ (with dark-baryon number 2) and $\X$ (with dark-baryon number 1) with masses $M_\X=M_\W/\sqrt{2}$. After confinement of $\Sp(\N)$ dynamics, two $\X$ can either form neutral mesons that decay to SM particles or charged mesons that decay into $\W$, leaving the $\W$ as the only DM candidate of the model.

In section \ref{Sommerfeld} we discussed how to compute the corrections to DM annihilation cross sections due to the exchange of light mediators (namely the Sommerfeld enhancement and bound state formation), and we focused on the relevant processes which occur in both cosmological evolution and indirect detection signals.

In section \ref{DMrelic} we computed the DM relic abundance by solving numerically the Boltzmann equations of $\W$ and $\X$ abundances at the perturbative level, and we also took into account the gauge confinement under some simplified assumptions. DM is accompanied by light, unstable glueballs, which can dilute the DM abundance and can be potentially probed through their coupling to the Higgs. We discussed the impact of the Sommerfeld enhancement and perturbative bound state formation on the relic abundance. All these results are summarized in Figure\fig{DMabundancefig}. 
Besides, we also show the predictions of Higgs and glueballs phenomenology in the right panel of figure\fig{Scalonfig}.

In section \ref{pheno1} we studied the signatures of this model. We calculated the scalon production cross section and the results are shown in the left panel of figure\fig{Scalonfig}. Besides, we also computed cross sections for direct and indirect detection and compared them with present experimental data and future prospects. We predicted a non standard indirect detection cross section, which turns out to be slightly suppressed in comparison to the usual WIMPs paradigm, and this is a characteristic and testable prediction of our model. The region in which the DM abundance 
is allowed by experimental bounds can be further explored by the future direct and indirect detection experiments, in particular the CTA.
The phenomenology for direct and indirect detection is summarized in figure\fig{DDfig}. 
 
In section \ref{sec:scale_inv} we gave a brief investigation of the effects of the quadratic parameter $M_\S^2$ on the mass spectrum as well as the phenomenological
results. Under non scale invariant conditions, we found that the relic density and indirect detection are roughly
the same as in the scale invariant case, while the direct detection cross section $\sigma_{\text{SI}}$ and the Higgs-glueballs mixing $\epsilon^2$ get suppressed. Besides, the theory becomes less  predictive due to more free parameters.

As we sometimes relied on approximations, various aspects of the model can be more precisely computed. For example, if $\Lambda_{\text{DC}} \gtrsim T_f$ (i.e. confinement occurs before DM thermal freeze-out), the computation of the relic abundance will be qualitatively different. Therefore, it can be interesting for a new work to study this case with more detailed computation. 

\subsubsection*{Acknowledgements} 
The work of JWW is supported by
the China Scholarship Council with Grant No. 201904910660. We thank Alessandro Strumia, Daniele Teresi, Luca Di Luzio, and Christian Gross for discussions.

\appendix

	\section{Feynman Rules: $\Sp(\N+2)$$\rightarrow$$\Sp(\N)$}
	\label{app:SpNFeyn}

The Feynman vertices with all momenta $p_i$ incoming are: 
\begin{itemize}
	\item[] \hspace{-1cm} 
	\begin{minipage}{0.2\textwidth}
		$\A(1)\A(2)\A(3):$
	\end{minipage}
	\begin{minipage}{0.5\textwidth}
		\vspace{-0.65cm}
		\beq 
		\ \g f^{a_1a_2a_3}_{\N} \[ 
		g_{\mu_1 \mu_2} (p_1^{\mu_3}-p_2^{\mu_3}) 
		- g_{\mu_1 \mu_3} (p_1^{\mu_2}-p_3^{\mu_2}) 
		+ g_{\mu_2 \mu_3} (p_2^{\mu_1}-p_3^{\mu_1})
		\] \nonumber
		\eeq
	\end{minipage}
     \vspace{-0.7cm}
     \item[]\hspace{-1cm}
	\begin{minipage}{0.2\textwidth}
		$\A(1)\A(2)\A(3)\A(4):$
	\end{minipage}
	\begin{minipage}{0.6\textwidth}
		\vspace{0.7cm}
		\centering
		\begin{align} 
		\hspace{-1.5cm}
		\ i \g^2 
		&\[
		f^{a_1a_3c}_{\N} f^{a_2a_4c}_{\N} 
		(g_{\mu_1 \mu_4} g_{\mu_2 \mu_3} - g_{\mu_1 \mu_2} g_{\mu_3 \mu_4}) \right. \nonumber \\
		& + f^{a_1a_2c}_{\N} f^{a_3a_4c}_{\N} 
		(g_{\mu_1 \mu_4} g_{\mu_2 \mu_3} - g_{\mu_1 \mu_3} g_{\mu_2 \mu_4}) \nonumber \\
		& \left. + f^{a_1a_4c}_{\N} f^{a_2a_3c}_{\N} 
		(g_{\mu_1 \mu_3} g_{\mu_2 \mu_4} - g_{\mu_1 \mu_2} g_{\mu_3 \mu_4}) \] \nonumber
		\end{align}
	\end{minipage}
 \vspace{-0.2cm}
	\item[] \hspace{-1cm} 
	\begin{minipage}{0.2\textwidth}
		$\A(1)\X(2)\X^{\dagger}(3):$
	\end{minipage}
	\begin{minipage}{0.5\textwidth}
			\vspace{-0.4cm}
		\begin{align}
		\ i \g (T^{a_1}_{\N})_{m_3 m_2} 
		&\[ 
		g_{\mu_1\mu_2} (p_2^{\mu_3}- p_1^{\mu_3}) 
		- g_{\mu_1\mu_3} (p_3^{\mu_2}- p_1^{\mu_2})  
		+ g_{\mu_2\mu_3} (p_3^{\mu_1}-p_2^{\mu_1})
		\] \nonumber 
		\end{align}
	\end{minipage}
\vspace{-0.7cm}
	\item[] \hspace{-1cm} 
	\begin{minipage}{0.2\textwidth}
		$\A(1)\A(2)\X(3)\X^{\dagger}(4):$
	\end{minipage}
	\begin{minipage}{0.6\textwidth}
		\vspace{0.7cm}
		\begin{align} 
		\ i \g^2 
		&\[ 
		(T^{a_1}_{\N} T^{a_2}_{\N})_{m_4 m_3} 
		\(g_{\mu_1\mu_3} g_{\mu_2\mu_4} 
		- g_{\mu_1\mu_2} g_{\mu_3\mu_4}\) \right. \nonumber \\
		& + (T^{a_2}_{\N} T^{a_1}_{\N})_{m_4 m_3}  \( 
		g_{\mu_1\mu_4} g_{\mu_2\mu_3}
		- g_{\mu_1\mu_2} g_{\mu_3\mu_4} \) \nonumber \\
		& \left. -i f^{a_1a_2c}_{\N} (T^{c}_{\N})_{m_4 m_3} \( g_{\mu_1\mu_4} g_{\mu_2\mu_3} 
		- g_{\mu_1\mu_3} g_{\mu_2\mu_4} \)
		\] \nonumber 
		\end{align}
	\end{minipage}
\vspace{0.3cm}
	\item[] \hspace{-1cm} 
	\begin{minipage}{0.2\textwidth}
		$\W(1)\W^{\dagger}(2)\Z(3):$
	\end{minipage}
	\begin{minipage}{0.5\textwidth}
		\vspace{-0.8cm}
		\beq 
		\ - i \g  \[ 
		g_{\mu_1 \mu_2} (p_1^{\mu_3}-p_2^{\mu_3}) 
		- g_{\mu_1 \mu_3} (p_1^{\mu_2}-p_3^{\mu_2}) 
		+ g_{\mu_2 \mu_3} (p_2^{\mu_1}-p_3^{\mu_1})
		\] \nonumber 
		\eeq
	\end{minipage}
\vspace{0.2cm}
	\item[] \hspace{-1cm}
	\begin{minipage}{0.2\textwidth}
		$\X(1)\X^{\dagger}(2)\Z(3):$
	\end{minipage}
	\begin{minipage}{0.5\textwidth}
		\vspace{-0.8cm}
		\beq 
		\ - i \frac{\g}{2}(\delta_{\N})_{m_1 m_2}  \[ 
		g_{\mu_1 \mu_2} (p_1^{\mu_3}-p_2^{\mu_3}) 
		- g_{\mu_1 \mu_3} (p_1^{\mu_2}-p_3^{\mu_2}) 
		+ g_{\mu_2 \mu_3} (p_2^{\mu_1}-p_3^{\mu_1})
		\] \nonumber 
		\eeq
	\end{minipage}
\vspace{0.2cm}
	\item[] \hspace{-1cm} 
	\begin{minipage}{0.2\textwidth}
		$\W(1)\W^{\dagger}(2)\Z(3)\Z(4):$
	\end{minipage}
	\begin{minipage}{0.7\textwidth}
		\vspace{-0.3cm}
		\beq 
		\hspace{-1.8cm}
		\ -i {\g^2} \( 
		2 g_{\mu_1\mu_2} g_{\mu_3\mu_4} - g_{\mu_1\mu_4} g_{\mu_2\mu_3}
		- g_{\mu_1\mu_3} g_{\mu_2\mu_4}
		\) \nonumber 
		\eeq
	\end{minipage}
	\item[] \hspace{-1cm}
	\begin{minipage}{0.2\textwidth}
		$\X(1)\X^{\dagger}(2)\Z(3)\Z(4):$
	\end{minipage}
	\begin{minipage}{0.7\textwidth}
		\vspace{-0.1cm}
		\beq 
		\ -i \frac{{\g^2}}{4}(\delta_{\N})_{m_1 m_2} \( 
		2 g_{\mu_1\mu_2} g_{\mu_3\mu_4} - g_{\mu_1\mu_4} g_{\mu_2\mu_3}
		- g_{\mu_1\mu_3} g_{\mu_2\mu_4}
		\) \nonumber 
		\eeq
	\end{minipage}
	\item[] \hspace{-1cm} 
	\begin{minipage}{0.2\textwidth}
		$\X(1)\X^{\dagger}(2)\Z(3)\A(4):$
	\end{minipage}
	\begin{minipage}{0.7\textwidth}
		\vspace{-0.2cm}
		\beq 
		\ -i \frac{\g^2}{2}(T_{\N}^{a_4})_{m_2 m_1} \( 
		2 g_{\mu_1\mu_2} g_{\mu_3\mu_4} - g_{\mu_1\mu_4} g_{\mu_2\mu_3}
		- g_{\mu_1\mu_3} g_{\mu_2\mu_4}
		\) \nonumber 
		\eeq
	\end{minipage}
	\item[] \hspace{-1cm} 
	\begin{minipage}{0.2\textwidth}
		$\s(1)\X(2)\X^{\dagger}(3):$
	\end{minipage}
	\begin{minipage}{0.2\textwidth}
		\vspace{-0.6cm}
		\beq 
		\hspace{0.6cm}
		\ 2 i \frac{M^2_\X}{w} (\delta_{\N})_{m_2m_3} g_{\mu_2\mu_3} \nonumber 
		\eeq
	\end{minipage}
	\item[] \hspace{-1cm} 
	\begin{minipage}{0.25\textwidth}
		$\s(1)\s(2)\X(3)\X^{\dagger}(4):$
	\end{minipage}
	\begin{minipage}{0.2\textwidth}
		\vspace{-0.65cm}
		\beq 
		\hspace{-0.2cm}
		\ 2 i \frac{M^2_\X}{w^2} (\delta_{\N})_{m_3m_4} g_{\mu_3\mu_4} \nonumber 
		\eeq
	\end{minipage}
	\item[] \hspace{-1cm}
	\begin{minipage}{0.2\textwidth}
		$\s(1)\W(2)\W^{\dagger}(3):$
	\end{minipage}
	\begin{minipage}{0.2\textwidth}
		\vspace{-0.65cm}
		\beq 
		\ 2 i \frac{M^2_\W}{w}  g_{\mu_2\mu_3} \nonumber 
		\eeq
	\end{minipage}
	\item[] \hspace{-1cm} 
	\begin{minipage}{0.2\textwidth}
		$\s(1)\s(2)\W(3)\W^{\dagger}(4):$
	\end{minipage}
	\begin{minipage}{0.2\textwidth}
		\vspace{-0.65cm}
		\beq 
		\ 2 i \frac{M^2_\W}{w^2} g_{\mu_3\mu_4} \nonumber 
		\eeq
	\end{minipage}
	\item[] \hspace{-1cm} 
	\begin{minipage}{0.2\textwidth}
		$\s(1)\Z(2)\Z(3):$
	\end{minipage}
	\begin{minipage}{0.2\textwidth}
		\vspace{-0.65cm}
		\beq 
		\ 2 i \frac{M^2_\Z}{w} g_{\mu_2\mu_3} \nonumber 
		\eeq
	\end{minipage}
	\item[] \hspace{-1cm} 
	\begin{minipage}{0.2\textwidth}
		$\s(1)\s(2)\Z(3)\Z(4):$
	\end{minipage}
	\begin{minipage}{0.2\textwidth}
		\vspace{-0.65cm}
		\beq 
		\ 2 i \frac{M^2_\Z}{w^2} g_{\mu_3\mu_4} \nonumber 
		\eeq
	\end{minipage}
	\item[] \hspace{-1cm} 
	\begin{minipage}{0.3\textwidth}
		$\W^{\dagger}(1)\X(2)\X(3)\Z(4):$
	\end{minipage}
	\begin{minipage}{0.3\textwidth}
		\vspace{-0.6cm}
		\beq 
		\hspace{-0.9cm} -\frac{3i\g^2}{2\sqrt{2}}\gamma_{\N}^{m_2m_3}(g_{\mu_1\mu_3}g_{\mu_2\mu_4}-g_{\mu_1\mu_2}g_{\mu_3\mu_4})  \nonumber 
		\eeq
	\end{minipage}
	\item[] \hspace{-1cm} 
	\begin{minipage}{0.2\textwidth}
		$\W(1)\W^{\dagger}(2)\X(3)\X^{\dagger}(4):$
	\end{minipage}
	\begin{minipage}{0.8\textwidth}
		\vspace{-0.2cm}
		\beq 
		\hspace{-2cm}
		\frac{i\g^2}{2}(\delta_{\N})_{m_3m_4}(2g_{\mu_1\mu_3}g_{\mu_2\mu_4}-g_{\mu_1\mu_4}g_{\mu_2\mu_3}-g_{\mu_1\mu_2}g_{\mu_3\mu_4})  \nonumber 
		\eeq
	\end{minipage}
	\item[] \hspace{-1cm} 
	\begin{minipage}{0.2\textwidth}
		$\W^{\dagger}(1)\X(2)\X(3):$
	\end{minipage}
	\begin{minipage}{0.8\textwidth}
		\vspace{-0.3cm}
		\beq 
		\hspace{0.6cm}
		-\frac{i\g}{\sqrt{2}}\gamma_{\N}^{m_2m_3}[g_{\mu_1\mu_2}(p_1^{\mu_3}-p_2^{\mu_3})-g_{\mu_3\mu_1}(p_1^{\mu_2}-p_3^{\mu_2})+g_{\mu_2\mu_3}(p_2^{\mu_1}-p_3^{\mu_1})]  \nonumber 
		\eeq
	\end{minipage}
    \vspace{-0.2cm}
	\item[] \hspace{-1cm} 
	\begin{minipage}{0.2\textwidth}
		$\A(1)\W^{\dagger}(2)\X(3)\X(4):$
	\end{minipage}
	\begin{minipage}{0.8\textwidth}
		\vspace{-0.1cm}
		\beq 
		\hspace{-1.6cm}
		\frac{i\g^2}{\sqrt{2}}(\gamma_{\N} T^{a_1}_{\N})_{m_3m_4}[2 g_{\mu_1\mu_2}g_{\mu_3\mu_4}-g_{\mu_1\mu_4}g_{\mu_2\mu_3}-g_{\mu_1\mu_3}g_{\mu_2\mu_4}]  \nonumber 
		\eeq
	\end{minipage}
\end{itemize}

\bibliographystyle{utphys}
\bibliography{cite_temp}

\providecommand{\href}[2]{#2}\begingroup\raggedright\begin{thebibliography}{10}

\bibitem{Buttazzo:2019iwr}
D.~Buttazzo, L.~Di~Luzio, G.~Landini, A.~Strumia, and D.~Teresi, ``{Dark Matter
  from self-dual gauge/Higgs dynamics},''
  \href{http://dx.doi.org/10.1007/JHEP10(2019)067}{{\em JHEP} {\bfseries 10}
  (2019) 067},
\href{http://arxiv.org/abs/1907.11228}{{\ttfamily arXiv:1907.11228 [hep-ph]}}.

\bibitem{Buttazzo:2019mvl}
D.~Buttazzo, L.~Di~Luzio, P.~Ghorbani, C.~Gross, G.~Landini, A.~Strumia,
  D.~Teresi, and J.-W. Wang, ``{Scalar gauge dynamics and Dark Matter},''
  \href{http://dx.doi.org/10.1007/JHEP01(2020)130}{{\em JHEP} {\bfseries 01}
  (2020) 130},
\href{http://arxiv.org/abs/1911.04502}{{\ttfamily arXiv:1911.04502 [hep-ph]}}.

\bibitem{Hambye:2009fg}
T.~Hambye and M.~H.~G. Tytgat, ``{Confined hidden vector dark matter},''
  \href{http://dx.doi.org/10.1016/j.physletb.2009.11.050}{{\em Phys. Lett.}
  {\bfseries B683} (2010) 39--41},
\href{http://arxiv.org/abs/0907.1007}{{\ttfamily arXiv:0907.1007 [hep-ph]}}.

\bibitem{Mitridate:2017izz}
A.~Mitridate, M.~Redi, J.~Smirnov, and A.~Strumia, ``{Cosmological Implications
  of Dark Matter Bound States},''
  \href{http://dx.doi.org/10.1088/1475-7516/2017/05/006}{{\em JCAP} {\bfseries
  1705} (2017) 006},
\href{http://arxiv.org/abs/1702.01141}{{\ttfamily arXiv:1702.01141 [hep-ph]}}.

\bibitem{Oncala:2018bvl}
R.~Oncala and K.~Petraki, ``{Dark matter bound states via emission of scalar
  mediators},'' \href{http://dx.doi.org/10.1007/JHEP01(2019)070}{{\em JHEP}
  {\bfseries 01} (2019) 070},
\href{http://arxiv.org/abs/1808.04854}{{\ttfamily arXiv:1808.04854 [hep-ph]}}.

\bibitem{Morselli:2017ree}
{\bfseries CTA Consortium} Collaboration, A.~Morselli, ``{The Dark Matter
  Programme of the Cherenkov Telescope Array},''
  \href{http://dx.doi.org/10.22323/1.301.0921}{{\em PoS} {\bfseries ICRC2017}
  (2018) 921}, \href{http://arxiv.org/abs/1709.01483}{{\ttfamily
  arXiv:1709.01483 [astro-ph.IM]}}.
[35,921(2017)].

\bibitem{Hambye:2013sna}
T.~Hambye and A.~Strumia, ``{Dynamical generation of the weak and Dark Matter
  scale},'' \href{http://dx.doi.org/10.1103/PhysRevD.88.055022}{{\em Phys.
  Rev.} {\bfseries D88} (2013) 055022},
\href{http://arxiv.org/abs/1306.2329}{{\ttfamily arXiv:1306.2329 [hep-ph]}}.

\bibitem{Hambye:2008bq}
T.~Hambye, ``{Hidden vector dark matter},''
  \href{http://dx.doi.org/10.1088/1126-6708/2009/01/028}{{\em JHEP} {\bfseries
  01} (2009) 028},
\href{http://arxiv.org/abs/0811.0172}{{\ttfamily arXiv:0811.0172 [hep-ph]}}.

\bibitem{Witten:1983tx}
E.~Witten, ``{Current Algebra, Baryons, and Quark Confinement},''
\href{http://dx.doi.org/10.1016/0550-3213(83)90064-0}{{\em Nucl. Phys.}
  {\bfseries B223} (1983) 433--444}.

\bibitem{andp.19314030302}
A.~Sommerfeld, ``Über die Beugung und Bremsung der Elektronen,''
  \href{http://dx.doi.org/10.1002/andp.19314030302}{{\em Annalen der Physik}
  {\bfseries 403} (1931) 257--330}.

\bibitem{Sakharov:1991pia}
A.~D. Sakharov, ``{Interaction of an Electron and Positron in Pair
  Production},'' \href{http://dx.doi.org/10.1070/PU1991v034n05ABEH002492}{{\em
  Sov.\ Phys.\ Usp.} {\bfseries 34} (1991) 375--377}.

\bibitem{Hisano:2006nn}
J.~Hisano, S.~Matsumoto, M.~Nagai, O.~Saito, and M.~Senami, ``{Non-perturbative
  effect on thermal relic abundance of dark matter},''
  \href{http://dx.doi.org/10.1016/j.physletb.2007.01.012}{{\em Phys. Lett.}
  {\bfseries B646} (2007) 34--38},
\href{http://arxiv.org/abs/hep-ph/0610249}{{\ttfamily arXiv:hep-ph/0610249
  [hep-ph]}}.

\bibitem{Cirelli:2007xd}
M.~Cirelli, A.~Strumia, and M.~Tamburini, ``{Cosmology and Astrophysics of
  Minimal Dark Matter},''
  \href{http://dx.doi.org/10.1016/j.nuclphysb.2007.07.023}{{\em Nucl. Phys.}
  {\bfseries B787} (2007) 152--175},
\href{http://arxiv.org/abs/0706.4071}{{\ttfamily arXiv:0706.4071 [hep-ph]}}.

\bibitem{Belotsky:2005dk}
K.~M. Belotsky, M.~{\relax Yu}. Khlopov, S.~V. Legonkov, and K.~I. Shibaev,
  ``{Effects of new long-range interaction: Recombination of relic heavy
  neutrinos and antineutrinos},'' {\em Grav. Cosmol.} {\bfseries 11} (2005)
  27--33,
\href{http://arxiv.org/abs/astro-ph/0504621}{{\ttfamily arXiv:astro-ph/0504621
  [astro-ph]}}.

\bibitem{Wise:2014jva}
M.~B. Wise and Y.~Zhang, ``{Stable Bound States of Asymmetric Dark Matter},''
  \href{http://dx.doi.org/10.1103/PhysRevD.90.055030,
  10.1103/PhysRevD.91.039907}{{\em Phys. Rev.} {\bfseries D90} (2014) 055030},
  \href{http://arxiv.org/abs/1407.4121}{{\ttfamily arXiv:1407.4121 [hep-ph]}}.
[Erratum: Phys. Rev.D91,no.3,039907(2015)].

\bibitem{vonHarling:2014kha}
B.~von Harling and K.~Petraki, ``{Bound-state formation for thermal relic dark
  matter and unitarity},''
  \href{http://dx.doi.org/10.1088/1475-7516/2014/12/033}{{\em JCAP} {\bfseries
  1412} (2014) 033},
\href{http://arxiv.org/abs/1407.7874}{{\ttfamily arXiv:1407.7874 [hep-ph]}}.

\bibitem{Petraki:2015hla}
K.~Petraki, M.~Postma, and M.~Wiechers, ``{Dark-matter bound states from
  Feynman diagrams},'' \href{http://dx.doi.org/10.1007/JHEP06(2015)128}{{\em
  JHEP} {\bfseries 06} (2015) 128},
\href{http://arxiv.org/abs/1505.00109}{{\ttfamily arXiv:1505.00109 [hep-ph]}}.

\bibitem{Ellis:2015vaa}
J.~Ellis, F.~Luo, and K.~A. Olive, ``{Gluino Coannihilation Revisited},''
  \href{http://dx.doi.org/10.1007/JHEP09(2015)127}{{\em JHEP} {\bfseries 09}
  (2015) 127},
\href{http://arxiv.org/abs/1503.07142}{{\ttfamily arXiv:1503.07142 [hep-ph]}}.

\bibitem{Asadi:2016ybp}
P.~Asadi, M.~Baumgart, P.~J. Fitzpatrick, E.~Krupczak, and T.~R. Slatyer,
  ``{Capture and Decay of Electroweak WIMPonium},''
  \href{http://dx.doi.org/10.1088/1475-7516/2017/02/005}{{\em JCAP} {\bfseries
  1702} (2017) 005},
\href{http://arxiv.org/abs/1610.07617}{{\ttfamily arXiv:1610.07617 [hep-ph]}}.

\bibitem{Liew:2016hqo}
S.~P. Liew and F.~Luo, ``{Effects of QCD bound states on dark matter relic
  abundance},'' \href{http://dx.doi.org/10.1007/JHEP02(2017)091}{{\em JHEP}
  {\bfseries 02} (2017) 091},
\href{http://arxiv.org/abs/1611.08133}{{\ttfamily arXiv:1611.08133 [hep-ph]}}.

\bibitem{Cirelli:2016rnw}
M.~Cirelli, P.~Panci, K.~Petraki, F.~Sala, and M.~Taoso, ``{Dark Matter's
  secret liaisons: phenomenology of a dark U(1) sector with bound states},''
  \href{http://dx.doi.org/10.1088/1475-7516/2017/05/036}{{\em JCAP} {\bfseries
  1705} (2017) 036},
\href{http://arxiv.org/abs/1612.07295}{{\ttfamily arXiv:1612.07295 [hep-ph]}}.

\bibitem{Morningstar:1999rf}
C.~J. Morningstar and M.~J. Peardon, ``{The Glueball spectrum from an
  anisotropic lattice study},''
  \href{http://dx.doi.org/10.1103/PhysRevD.60.034509}{{\em Phys.\ Rev.\ D}
  {\bfseries 60} (1999) 034509},
  \href{http://arxiv.org/abs/hep-lat/9901004}{{\ttfamily
  arXiv:hep-lat/9901004}}.

\bibitem{Mitridate:2017oky}
A.~Mitridate, M.~Redi, J.~Smirnov, and A.~Strumia, ``{Dark Matter as a weakly
  coupled Dark Baryon},'' \href{http://dx.doi.org/10.1007/JHEP10(2017)210}{{\em
  JHEP} {\bfseries 10} (2017) 210},
\href{http://arxiv.org/abs/1707.05380}{{\ttfamily arXiv:1707.05380 [hep-ph]}}.

\bibitem{Alekhin:2015byh}
S.~Alekhin {\em et~al.}, ``{A facility to Search for Hidden Particles at the
  CERN SPS: the SHiP physics case},''
  \href{http://dx.doi.org/10.1088/0034-4885/79/12/124201}{{\em Rept. Prog.
  Phys.} {\bfseries 79} (2016) 124201},
\href{http://arxiv.org/abs/1504.04855}{{\ttfamily arXiv:1504.04855 [hep-ph]}}.

\bibitem{Aprile:2018dbl}
{\bfseries XENON} Collaboration, E.~Aprile {\em et~al.}, ``{Dark Matter Search
  Results from a One Ton-Year Exposure of XENON1T},''
  \href{http://dx.doi.org/10.1103/PhysRevLett.121.111302}{{\em Phys. Rev.
  Lett.} {\bfseries 121} (2018) 111302},
\href{http://arxiv.org/abs/1805.12562}{{\ttfamily arXiv:1805.12562
  [astro-ph.CO]}}.

\bibitem{Ackermann:2015zua}
{\bfseries Fermi-LAT} Collaboration, M.~Ackermann {\em et~al.}, ``{Searching
  for Dark Matter Annihilation from Milky Way Dwarf Spheroidal Galaxies with
  Six Years of Fermi Large Area Telescope Data},''
  \href{http://dx.doi.org/10.1103/PhysRevLett.115.231301}{{\em Phys. Rev.
  Lett.} {\bfseries 115} (2015) 231301},
\href{http://arxiv.org/abs/1503.02641}{{\ttfamily arXiv:1503.02641
  [astro-ph.HE]}}.

\bibitem{Abdallah:2016ygi}
{\bfseries H.E.S.S.} Collaboration, H.~Abdallah {\em et~al.}, ``{Search for
  dark matter annihilations towards the inner Galactic halo from 10 years of
  observations with H.E.S.S},''
  \href{http://dx.doi.org/10.1103/PhysRevLett.117.111301}{{\em Phys. Rev.
  Lett.} {\bfseries 117} (2016) 111301},
\href{http://arxiv.org/abs/1607.08142}{{\ttfamily arXiv:1607.08142
  [astro-ph.HE]}}.

\bibitem{Buttazzo:2015bka}
D.~Buttazzo, F.~Sala, and A.~Tesi, ``{Singlet-like Higgs bosons at present and
  future colliders},'' \href{http://dx.doi.org/10.1007/JHEP11(2015)158}{{\em
  JHEP} {\bfseries 11} (2015) 158},
\href{http://arxiv.org/abs/1505.05488}{{\ttfamily arXiv:1505.05488 [hep-ph]}}.

\bibitem{Buttazzo:2018qqp}
D.~Buttazzo, D.~Redigolo, F.~Sala, and A.~Tesi, ``{Fusing Vectors into Scalars
  at High Energy Lepton Colliders},''
  \href{http://dx.doi.org/10.1007/JHEP11(2018)144}{{\em JHEP} {\bfseries 11}
  (2018) 144},
\href{http://arxiv.org/abs/1807.04743}{{\ttfamily arXiv:1807.04743 [hep-ph]}}.

\end{thebibliography}\endgroup

\end{document}